\documentclass{jfm}
\usepackage{graphicx}
\usepackage{multirow}
\usepackage{tabularx}
\usepackage{array}
\usepackage{epstopdf, epsfig}
\usepackage{amssymb}
\usepackage{amsmath}
\usepackage[dvipsnames]{xcolor}
\usepackage[]{lineno}
\usepackage{caption}
\usepackage{subcaption}

\shorttitle{Linear models of strip-type roughness}
\shortauthor{D. Lasagna, G. Zampino and B. Ganapathisubramani}

\title{Linear models of strip-type roughness}

\author{D. Lasagna\aff{1}\corresp{\email{davide.lasagna@soton.ac.uk}},
  G. Zampino\aff{2}
 \and B. Ganapathisubramani\aff{1}}

\affiliation{\aff{1}Aeronautics and Astronautics, Faculty of Engineering and Physical Sciences, University of Southampton, Hampshire, SO17 1BJ, UK \aff{2} {FLOW, Faculty of Engineering Mechanics, KTH Royal Institute of Technology
Stockholm, Sweden}}

\begin{document}
% \linenumbers
\maketitle

\begin{abstract}
Prandtl's secondary flows of the second kind generated by laterally-varying roughness are studied using the linearised Reynolds-Averaged Navier-Stokes approach proposed in \citet{zampino2022}. The momentum equations are coupled to the Spalart-Allmaras model while the roughness is captured by adapting established strategies for homogeneous roughness to heterogeneous surfaces. Linearisation of the governing equations yields a framework that enables a rapid exploration of the parameter space associated with heterogeneous surfaces, {in the limiting case of small spanwise variations of the roughness properties}. Channel flow is considered, with longitudinal high and low roughness strips arranged symmetrically. By varying the strip width, it is found that linear mechanisms play a dominant role in determining the size and intensity of secondary flows. In this setting, secondary flows may be interpreted as the time-averaged output response of the turbulent mean flow subjected to a steady forcing produced by the wall heterogeneity. In fact, the linear model predicts that secondary flows are most intense when the strip width is about 0.7 times the half-channel height, in excellent agreement with available data. Furthermore, a unified framework to analyse combinations of heterogeneous roughness properties and laterally-varying topographies, common in applications, is discussed. {Noting that the framework assumes small spanwise variations of the surface properties, two separate secondary-flow inducing source mechanisms are identified}, i.e. the lateral variation of the virtual origin from which the turbulent structure develops and the lateral variation of the streamwise velocity slip, capturing the acceleration/deceleration perceived by the bulk flow over troughs and crests of non-planar topographies.

% In engineering applications, the surfaces are not smooth even the ridges. The combination of strip-type and ridge-type roughness heterogeneity have been investigated by exploiting the linearity of the equations. Due to the difference in the boundary conditions used to model the roughness heterogeneity, at the same induced virtual origin, the present analysis demonstrated that effect of strips on the strength of secondary flows is dominant and the flow topology observed above the combined roughess heterogeneity changes in the flow direction, in agreement with the dominant roughness heterogeneity. 

\end{abstract}

\begin{keywords}
Linearised RANS equations, secondary flows, strip-type roughness
\end{keywords}
\section{Introduction}
Prandtl's secondary flows of the second kind \citep{prandtl1952} emerge when a turbulent flow develops over an heterogeneous surface with a lateral variation of its properties. Two equivalent standpoints, based on the analysis of the Reynolds-averaged equations, explain the formation of such currents. One standpoint considers secondary currents as the product of the imbalance between production and dissipation of turbulent kinetic energy induced by the roughness heterogeneity \citep{hinze1973}, whereby turbulence-rich fluid is advected towards low-turbulence regions. The second standpoint considers the streamwise vorticity balance \citep{perkins1970}, whereby cross-stream gradients of the Reynolds stresses arising from the cross-stream velocity components induce a turbulent torque that acts as a source term in the streamwise vorticity equation \citep{castro2024}. Overall, such mechanisms produce large-scale counter-rotating longitudinal rolls appearing in the time-averaged wall-bounded flow. The associated upwelling and downwelling motions produced by the rolls induce a lateral distortion of the boundary layer height \citep{barros2014}, together with alternating high and low streamwise momentum regions \citep{mejia2013,willigham2014, anderson2015}, arranged analogously to the classical roll-streak pattern in shear flows \citep{brandt2014}.

Secondary flows are commonly observed in many industrial and environmental applications, where surfaces are either characterised by lateral variations of the topography, i.e. the elevation, or of the friction, e.g. by means of varying roughness properties. These two types of heterogeneity have been idealised in the literature as ridge-type and strip-type roughness configurations, respectively. The first type consists of longitudinal ribs located on a smooth, planar surface having rectangular or more complex cross-sections \citep{goldstein1998, Hwanglee2018, zampiron2020, castro2020, long2023, Zampino2023, zhdanov2024influence}, or alternatively smooth sinusoidal modulations of the wall \citep{Wang2006, vidal2018}. The second type, the focus of this work, consists of alternating longitudinal strips of high and low roughness. Secondary motions over such an arrangement have been extensively characterised experimentally \citep{bai2018, wangsawijaya2020, wangsawijaya2022, Frohnapfel_von_Deyn_Yang_Neuhauser_Stroh_Gatti_2024}, and in numerical simulations \citep{willigham2014, anderson2015, chung2018, stroh2019, forooghi, Neuhauser_Schafer_Gatti_Frohnapfel_2022, schafer2022}.

% have been the focus of a number of studies since they influence wall-normal transport properties, heat transfer rates and the performance of aerodynamic surfaces (see \citet{mejia2013, barros2014, vanderwel2015, Hwanglee2018, medjnoun_vanderwel_ganapathisubramani_2018, chung2018, meyers2019, stroh2019, medjnoun_vanderwel_ganapathisubramani_2020,vidal2018, castro2020, zampiron2020, stroh2020, forooghi, wangsawijaya2020, zhdanov2024influence} among many others). \citep{von2022drag}

% or by the presence of super-hydrophobic features \citep{turk2014,anderson2015,stroh2016}. 

Despite the burgeoning interest in these flows and the intense examination of their characteristics, {there is a number of aspects} clearly documented in the literature for which a physics-based, mechanistic model is not available. The first aspect is related to the marked dependence of the size and intensity of secondary flows on one or more spanwise length scales characterising the surface heterogeneity \citep{wangsawijaya2020, yang2017, medjnoun_vanderwel_ganapathisubramani_2018}. For strip-type roughness this length scale is usually expressed by the width $S$ of the strips. Consensus has emerged on the existence of three separate regimes as $S$ varies in relation with the average boundary layer thickness $\delta$ \citep{chung2018}. When the strip width is much smaller than the boundary layer thickness,  $S\ll \delta$, secondary flows are confined to the vicinity of the surface and do not strongly influence the outer region. Conversely, when $S\gg\delta$ secondary flows are localised in regions where the surface properties vary more rapidly, and wide areas of local flow homogeneity are observed away from such regions \citep{wangsawijaya2020}. When $S \approx \delta$, the secondary flows are most intense and can significantly influence the flow structure. Nevertheless, a model that captures the nature of these regimes and identifies boundaries between them is not available at present. In addition, most studies have considered strips of equal width, but the width ratio between high and low roughness strips is certainly important, as it is for rectangular ridges where the ratio of recessed and elevated area influences the flow structure \citep{medjnoun_vanderwel_ganapathisubramani_2020, zampino2022}.

A second aspect is related to the occurrence of {the so-called tertiary flows. These are weaker longitudinal roll structures adjacent to the dominant rolls} often associated to a reversal of the vertical flow direction at the centre of the high (or low) roughness strip, or at the centre of the ridge (or trough) \citep{vanderwel2015}. Tertiary flows are commonly observed over surfaces with longitudinal ridges (e.g. \citet{medjnoun_vanderwel_ganapathisubramani_2020}), especially when the width of the troughs or of the ridges is large enough to accommodate multiple streamwise vortices next to each other. For heterogeneous rough surfaces, however, tertiary flows have not been observed. In fact, for wide strips, cross-stream motions have been observed to be mostly confined in a roughly square region around the transition between the strips. This applies to both boundary layer \citep{wangsawijaya2020} and channel flows \citep{chung2018, Neuhauser_Schafer_Gatti_Frohnapfel_2022}. One explanation may be that tertiary flows over roughness strips might be difficult to discern in the mean flows obtained from experiments or simulations, especially when instantaneous structures meander quite significantly in the longitudinal direction \citep{zampiron2021, kevin2019}, smearing weak cross-stream motions. \citet{Neuhauser_Schafer_Gatti_Frohnapfel_2022} hypothesised that the boundary conditions utilised in numerical simulations to capture the roughness effect may also play a role, although this {hypothesis} does not appear to explain why tertiary flows are not seen in experiments. 
% It has to be pointed out, though, that all the above-mentioned works examined strip configurations with duty cycle $DC=0.5$, and did not examined asymmetric configurations. 

A third aspect that still lacks a robust mechanistic explanation is motivated by features of realistic surfaces in engineering and natural applications, whereby lateral changes of the roughness height are almost invariably accompanied by a lateral change in the elevation \citep{stroh2019, schafer2022}. Decoupling these two effects may be easier in numerical simulations where the roughness heterogeneity is modelled by suitable spanwise heterogeneous boundary conditions applied to an otherwise planar boundary of the numerical domain \citep{chung2018, Neuhauser_Schafer_Gatti_Frohnapfel_2022}, but requires care when setting up experiments with, e.g. sandpaper strips or in roughness-resolving numerical simulations \citep{Frohnapfel_von_Deyn_Yang_Neuhauser_Stroh_Gatti_2024}. 
% The study of the ridge- and strip-type roughness combination takes form from the need to better describe the real features of the industrial surfaces. As largely discussed by the any change of the roughness height is "accompanied by a change in the surface height" because the realistic roughness is characterised by a mean roughness height respect to a reference smooth surface.  
 % or \emph{vice-versa} analysing their combination,
% In addition, the systematic analysis of the combination of the roughness heterogeneity would help to develop a strategy to amplify or diminish the size and strength of the secondary flows. 
% In engineering applications, the surfaces are not smooth even the ridges.  
One explicit attempt to study the coupling between these two effects was carried out by \citet{stroh2019} and then later by \citet{schafer2022} who performed a series of direct numerical simulations over surfaces characterised by alternating rough and smooth regions. In their paper, \citet{stroh2019} completely resolved the surface roughness using an immersed boundary method and studied three different configurations: the mean roughness height is (i) lower, (ii) equal to, and (iii) higher than the elevation of the smooth surface. The authors observed a change in the flow organisation moving from case (i) to (iii) and vice-versa. This behaviour was not reproduced in more recent roughness-resolving simulations \citep{Frohnapfel_von_Deyn_Yang_Neuhauser_Stroh_Gatti_2024}, which was attributed to the importance of the strip width, relative to the roughness height. 

One last aspect for which a model does not seem to be available concerns the relation between naturally-occurring Very-Large-Scale-Motions (VLSMs), populating the log-layer over homogeneous surfaces, and secondary flows \citep{chung2018, Lee_Sung_Adrian_2019, wangsawijaya2022}. It has been speculated that secondary flows may be interpreted as VLSMs locked in place by the surface heterogeneity, given some similarity in their features. This can readily explain why secondary flows are most intense when $S \approx \delta$, because the strip width is commensurate with the spanwise length scale of such motions. Evidence shows that VLSMs and secondary flows do indeed coexist and do interact to a significant extent, since energy from the former appear to leak into the latter \citep{zampiron2020, wangsawijaya2020}. However, the specific mechanism for which large scale structures residing in the outer layer should be locked in place so effectively by the roughness heterogeneity is not fully clear.

In  recent work \citep{zampino2022}, we developed a predictive framework to understand how far can linear mechanisms go in explaining these aspects, focusing on ridge-type roughness. The framework {originates} from the long line of work that relies on the Reynolds-Averaged Navier-Stokes equations, augmented with a turbulent viscosity model and linearised about the turbulent mean, to explain the structure of {smooth-wall turbulence (see \citet{delalamo2006, pujals2009, hwangcossu2010, McKEON_SHARMA_2010} and references therein), or as a systematic tool to investigate flow control strategies \citep{Moarref.2012, Luhar.2015} and patterned surfaces \citep{Chavarin.2020, Ran.2020}}. Differently to previous efforts \citep{meyers2019}, the framework utilises the Spalart-Allmaras equation \citep{spalart1994} to capture turbulent viscosity transport phenomena in combination with the nonlinear Quadratic Constitutive Relation (QCR, \citet{spalart2000}) to model the anisotropy of the Reynolds stress tensor, required to produce secondary motions (see \citet{speziale1982, speziale1991, bottaro2006}). {It also assumes that spanwise variations of the surface topography are infinitesimally small. This allows the mean response of the turbulent flow to be obtained using linear equations where different spanwise wavenumber components have decoupled. The relevance of the assumption for surfaces with finite-amplitude topographies remains to be examined, although recent work on rectangular ridges \citep{castro2024} suggests that this approximation may only be acceptable for ridges with moderate height. This, in turn, indicates that the mean response of the turbulent flow to a perturbation of the surface topography may be quite nonlinear. Nevertheless,} one first advantage of the linear framework is its computationally {efficiency}. It thus enables the vast parameter space characterising heterogeneous surfaces to be explored rapidly, for instance to unravel the effect of ridge geometry \citep{Zampino2023}. A second key advantage of the framework is that it provides a perspective of secondary motions as being the output response of the turbulent mean subjected to a steady perturbation produced by the surface heterogeneity. 

% Linear input-output analysis of the Reynolds-Averaged Navier-Stokes equations has demonstrated to be a powerful tool to explain the structure of wall-bounded turbulence \citep{hwangcossu2010, McKEON_SHARMA_2010}, and in \citet{zampino2022}, a first step towards bringing such tools to bear on the problem of characterising the nature of secondary motions produced by the surface heterogeneity was made.

In this paper, we bring the same framework to bear on the problem of strip-type roughness. {We assume that the spanwise variation of the roughness is small, so that linear equations governing the response of the flow can be obtained.} The effect of the surface roughness is introduced following well-established modelling strategies for rough walls \citep{Aupoix2007, prakash2020}. {Briefly}, such strategies consist of modifying the virtual origin from which the turbulent flow develops, in order to obtain the desired shift of the logarithmic velocity profile. 
The overall aim of this paper is to characterise the formation and structure of secondary flows developing above strip-type roughness by means of the proposed linear framework. This will allow fundamental insight into the linear mechanisms that control such flows to be generated. We apply the proposed linear framework to flows in channels, and examine the role that the surface arrangement plays on: (i) the strength of secondary motions as a function of the strip width, identifying the three regimes discussed in \citet{chung2018}, (ii) the occurrence of tertiary flows as the {relative width of the high and low roughness strips} is varied, (iii) the structure of low- and high-momentum pathways and finally (iv) the combination of roughness and surface elevation effects.

The modelling framework, and its extension to rough surfaces, is presented in section \ref{sec:methodology}. Results are then reported in section \ref{sec:strips}. In section \ref{sec:combination}, the framework is generalised to more complex surface heterogeneities, combining the effects of roughness and surface elevation. Finally, conclusions are summarised in section \ref{sec:conclusions}.     

\section{Methodology}\label{sec:methodology}
\subsection{Governing equations}
\label{sec:governing_equation}
The incompressible flow of a fluid with kinematic viscosity $\nu$ and density $\rho$ is considered in a pressure driven channel with half-height $h$ and subjected to a streamwise pressure gradient $\Pi$. The friction velocity $u_\tau=\sqrt{\tau_w/\rho}$, with $\tau_w = h \Pi$ the mean friction, yields the friction Reynolds number $\Rey_\tau=u_\tau h/\nu$. Index notation is used for the Cartesian coordinates $x_i$ and velocities components $u_i$. {Quantities are generally normalised by $h$ and $u_\tau$. The superscript $(\cdot)^+$ is omitted in the following to reduce clutter, unless necessary to identify a length scaled by the viscous length.}
% but we occasionally use this superscript and the wall-normal coordinate $y$, starting from zero, when necessary.
\begin{figure}
    \centering
    \includegraphics[width=0.82\textwidth]{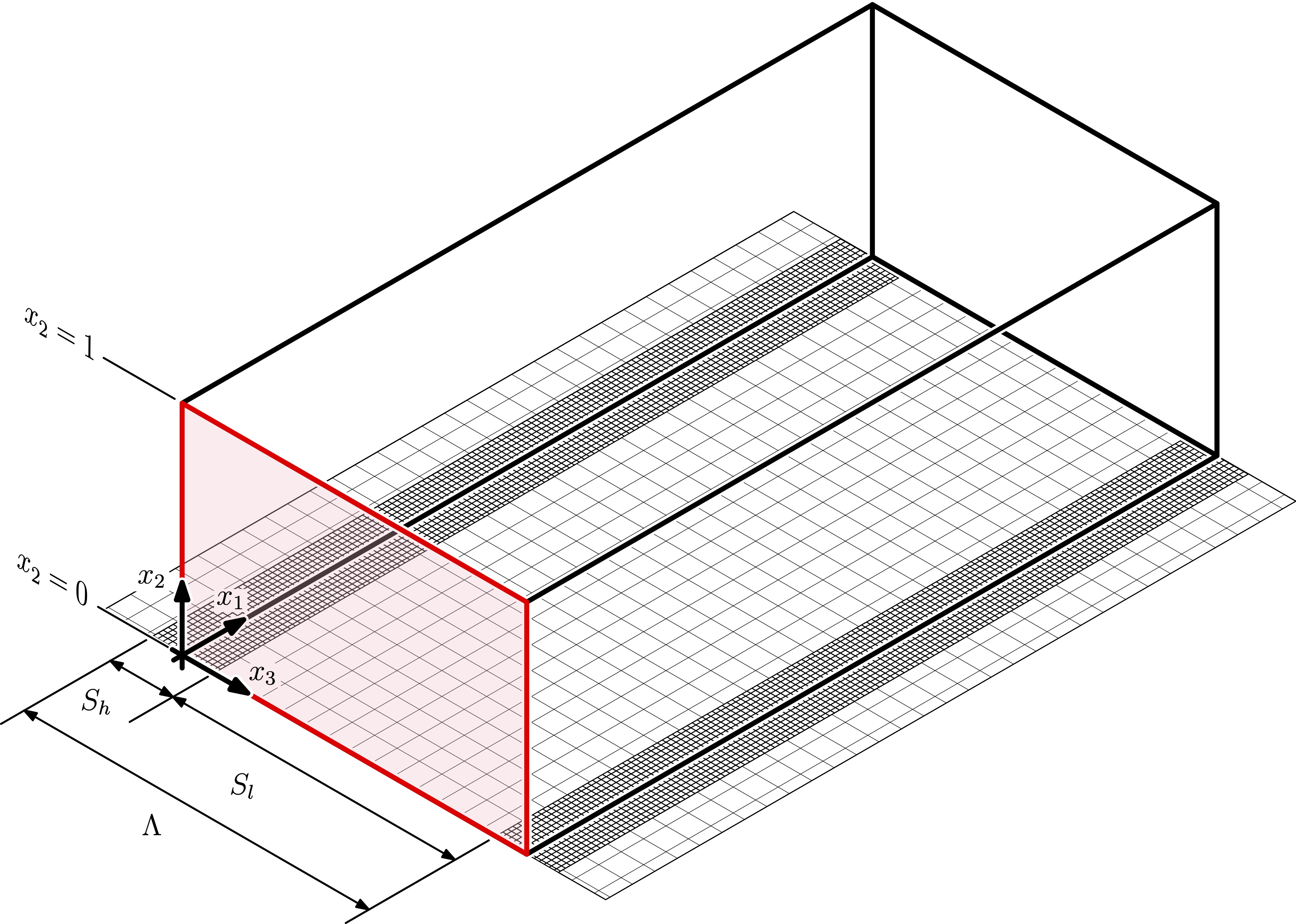}
    \caption{Sketch of the bottom half of the channel with high and low roughness strips, of width $S_h$ and $S_l$ respectively, placed symmetrically on the bottom and upper (not shown) walls and aligned with the streamwise direction $x_1$. This pattern repeats laterally with period \mbox{$\Lambda = S_h+S_l$}. These dimensions are normalised with the channel half-height $h$. The coordinate system is anchored at the bottom plane, at the centre of one of the high roughness strips. Owing to the symmetry of the strip arrangement on the two walls, the shaded red area defines the fundamental repeating flow unit in which the flow structure is visualised later in the paper.}
    \label{fig:sketch}
\end{figure}
The channel walls are covered by alternating strips of high and low roughness having width $S_h$ and $S_l$, respectively, as shown in figure \ref{fig:sketch}. The strips are streamwise-aligned and are placed symmetrically on the two walls. The pattern repeats with spanwise periodicity $\Lambda = S_h + S_l$, the fundamental length scale. We also introduce the duty cycle $DC = S_h / \Lambda$ to characterise the relative width of the strips, and refer to $S$ as the strip width when $S_h = S_l$, i.e.~ for $DC=0.5$.

  % The reference pressure is $p_{ref}=\rho u_\tau^2$ and the dimensionless pressure gradient [FIX] $\partial \bar{p}/\partial x_i = \delta_{i1}$, with $\delta_{ij}$ being the Kronocker delta, is the forcing term of the governing equations. 
  
The continuity and momentum equations are Reynolds averaged and made non-dimensional using $h$, $u_\tau$ and $\rho$. Average and fluctuation quantities are denoted by an overbar and a prime. For streamwise-aligned strips, we assume a streamwise-independent time-averaged flow, i.e. $\partial(\cdot)/\partial x_1 \equiv 0$, which filters out the meandering of secondary currents \citep{zampiron2020}. As a result, the mean pressure can be eliminated by considering the mean streamwise vorticity equation and introducing the streamfunction $\overline{\psi}$, satisfying $\nabla^2 \overline{\psi}=\overline{\omega}_1$ with 
\begin{equation}
    \overline{\omega}_1= {\frac{\partial \overline{u}_3}{\partial x_2}-\frac{\partial \overline{u}_2}{\partial x_3}}
    \label{omegadefinition}
\end{equation} 
the mean streamwise vorticity. {The cross-stream velocity components are $\overline{u}_2=-{\partial \overline{\psi}}/{\partial x_3}$ and $\overline{u}_3={\partial \overline{\psi}}/{\partial x_2}$.} The Reynolds-averaged equations for the streamwise momentum and the streamfunction are then
\begin{subeqnarray}
    \displaystyle
  \frac{\partial \overline{\psi}}{ \partial x_2} \displaystyle \frac{\partial \overline{u}_1}{ \partial x_3} - && \displaystyle \frac{\partial \overline{\psi}}{ \partial x_3} \displaystyle \frac{\partial \overline{u}_1}{ \partial x_2} \!= \!1\!+\!\frac{1}{\Rey_\tau} \!\left( \displaystyle \frac{\partial^2 \overline{u}_1}{\partial x_2^2} \!+ \! \displaystyle \frac{\partial^2 \overline{u}_1}{\partial x_3^2}\!\right) \!+\! \displaystyle \frac{\partial \tau_{12}}{\partial x_2} \!+ \! \displaystyle \frac{\partial \tau_{13}}{\partial x_3}, \label{systemRANSa}\\
\displaystyle
  \nonumber  \displaystyle \frac{\partial^2}{\partial x_2 \partial x_3} \! && \!\left[ \! \left(\displaystyle \frac{\partial \overline{\psi}}{\partial x_2}\right)^2 -\left( \displaystyle \frac{\partial \overline{\psi}}{\partial x_3}\right)^2 \! \right] \! + \!\left( \displaystyle \frac{\partial^2}{\partial x_3^2} \!- \! \displaystyle \frac{\partial^2}{\partial x_2^2}\right)\!\displaystyle  \frac{\partial \overline{\psi}}{\partial x_2} \! \displaystyle \frac{\partial \overline{\psi}}{\partial x_3} = \,\;\,\;\, \\
  \displaystyle  \,\;\,\;\;\; \frac{1}{\Rey_\tau} \!&& \!\left( \displaystyle \frac{\partial^2}{\partial x_2^2} \!+\! \displaystyle \frac{\partial^2}{\partial x_3^2} \!\right)^2 \overline{\psi} 
  \!+\! \displaystyle \frac{\partial^2}{\partial x_2 \partial x_3} \left( \tau_{33}\!-\!\tau_{22}\right) \!+\! \left( \displaystyle \frac{\partial^2}{\partial x_2^2} \!-\! \displaystyle \frac{\partial^2}{\partial x_3^2} \!\right) \tau_{23},
  \label{systemRANSb}
\end{subeqnarray}
where $\tau_{ij}= - \overline{u_i'u_j'}$ is the Reynolds stress tensor.

% In order to account for the change of the skin friction produced by the surface roughness, 
% \red{The streamwise velocity profile over rough surfaces is commonly written as 
% \begin{equation}
%     u_1^+=\frac{1}{k}\ln{\left( x_2^+\right)}+\Delta u^+=\frac{1}{k}\ln{\left( \frac{x_2^++d_s^+}{d_s^+}\right)}
%     \label{eq:def_log}
% \end{equation}
% where the virtual origin $d_s^+$, with the subscript denoting strip-type roughness, is defined as the distance from the wall where the streamwise velocity on the smooth surface is equal to the velocity deficit $\Delta u^+$. \red{In a channel flow we define $\Delta u^+=u_1(x_2=0)-u_{1_{smooth}}(x_2=0)$. [FIX: explain]} A different interpretation of the virtual origin was proposed by \citet{rotta1962} who observed that the roughness affects the development of the viscous sublayer. The law of the wall is therefore valid when the reference zero-velocity plane is shifted beneath the physical surface by $d_s$. }

\subsection{Turbulence modelling} \label{sec:turbulence modelling}
% The term at the left-hand side of \eqref{OSfinal} is analogous to the off-diagonal coupling operator in the Orr-Sommerfeld-Squire linearised equations \citep{schmid2000stability} and it is the only coupling term explicitly appearing in the system \eqref{eq:linear_equations}. 
When the linear Boussinesq hypothesis is used to express the deviatoric component of the Reynolds stresses as a function of the mean velocity gradients, namely 
\begin{equation}
    \tau_{ij}^{L} = 2 \nu_t S_{ij},
\end{equation}
where $\nu_t$ is the turbulent eddy viscosity and $S_{ij}$ is the {symmetric component of the} mean velocity gradient tensor
\begin{equation}
    S_{ij}=\frac{1}{2} \left(\frac{\partial \overline{u}_i}{\partial x_j} + \frac{\partial \overline{u}_j}{\partial x_i}\right),
\end{equation}
% the Reynolds stresses in the streamfunction equation {(\ref{psiequation}b)} decouple from the streamwise momentum equation {(\ref{psiequation}a)}. As demonstrated by \citet{perkins1970}, when the Boussinesq hypothesis is utilised the
the Reynolds stresses in {(\ref{systemRANSb})b} do not depend on the streamwise velocity. Then, the streamfunction equation decouples from the streamwise momentum equation and its solution is trivially $\overline{\psi}\equiv0$, i.e. no secondary flows are generated.

As extensively discussed in the literature (see e.g. \citet{perkins1970, speziale1982, bottaro2006}), a nonlinear stress model is needed to predict Prandlt's secondary flows of the second kind, produced by spatial gradients of the anisotropy of the Reynolds stresses. Several approaches have been proposed in literature (e.g. \citet{speziale1982,liencubic}). Here we utilise the Quadratic Constitutive Relation (QCR) nonlinear model presented in \citet{spalart2000}, whereby the deviatoric component of the Reynolds stresses becomes 
\begin{equation}\label{eq:taunonlinear}
     \tau_{ij}^{Q}=\tau_{ij}^{L}-c_{r1}\left[ O_{ik}\tau_{jk}^{L}+O_{jk}\tau_{ik}^{L}\right],
\end{equation} 
where $O_{ij}$ is the normalised rotation tensor defined as
\begin{equation}
O_{ij}  = {2W_{ij}} / {\sqrt{\displaystyle  {\frac{\partial \overline{u}_m}{\partial x_n} \frac{\partial \overline{u}_m}{\partial x_n}}}}, 
% \quad \mathrm{with}\quad
% W_{ij} = \frac{1}{2}\left( \frac{\partial \overline{u}_i}{\partial x_j}-\frac{\partial \overline{u}_j}{\partial x_i}\right),
\label{oijdefinition}
\end{equation}
and $W_{ij}$ is the anti-symmetric part of the velocity gradient tensor, {with $m$ and $n$ being summation indices}. The QCR model depends on a tuning single constant, whose value $c_{r1}=0.3$ was calibrated to match the anisotropy of the outer region of wall-bounded turbulent flows in \citet{spalart2000}. The default value is used throughout the paper.

{To close the momentum equations, a model for the eddy viscosity $\nu_t$ is necessary. Previous studies that have utilised the linearised Navier-Stokes equations have adopted analytical eddy viscosity profiles to analyse smooth-wall turbulent flows (see \citet{delalamo2006, pujals2009, hwangcossu2010, morra_2019} among others). Here, a complete transport model is preferred over such analytical ansatzs, as it is not clear a-priori how the eddy viscosity field should change when the mean flow structure is significantly distorted by secondary currents, or when roughness effects are important. For this purpose, the Spalart-Allmaras (SA) turbulence model \citep{spalart1994} is employed in this work}. The SA model is preferred here over other commonly employed two-equation models because it can be linearised relatively easily. In addition, the SA model was developed for attached shear flows, hence it should provide satisfactory predictions for the present case. The steady SA model defines a transport equation for the modified eddy viscosity $\tilde{\nu}$, normalised with $u_\tau$ and $h$. {This quantity is related to the turbulent viscosity by the relation 
\begin{equation}
\nu_t=\tilde{\nu} f_{v1},
\end{equation}
where $f_{v1}={\chi^3}/({\chi^3+c_{v1}^3})$ with $\chi= \Rey_\tau \tilde{\nu}$ and $c_{v1}$ a tuning constant. The modified eddy viscosity coincides with the turbulent viscosity away from the wall. The term $f_{v1}$ ensures the correct decay of the turbulent viscosity in the viscous sublayer \citep{spalart1994, mellor1968}, although $\tilde{\nu}$ behaves linearly in the log layer down to the wall, which is advantageous for numerical reasons. The transport equation is}
\begin{equation}
        \overline{u}_i \frac{\partial \tilde{\nu}}{\partial x_i} =c_{b1} \tilde{\mathcal{S}} \tilde{\nu}+\frac{1}{\sigma}\left\{ \frac{\partial }{\partial x_j}\left[ \left( \frac{1}{\Rey_\tau}+\tilde{\nu}\right) \frac{\partial \tilde{\nu} }{\partial x_j }\right] + c_{b2} \frac{\partial \tilde{\nu}}{ \partial x_j}\frac{\partial \tilde{\nu}}{ \partial x_j}  \right\}-c_{w1} f_w \left( \frac{\tilde{\nu}}{d}\right)^2,
    \label{SAequation1}
\end{equation}
where the terms model, respectively, advection, production, diffusion and destruction. {In the production term, the quantity $\tilde{\mathcal{S}}$ is defined as
\begin{equation}
 \tilde{\mathcal{S}}=\sqrt{2W_{ij} W_{ij}}+ \displaystyle \frac{\tilde{\nu}}{\kappa^2 d^2} f_{v2} \quad \mathrm{with}  \quad f_{v2}= 1- \displaystyle \frac{\chi}{1+ \chi f_{v1}}.
\end{equation}
% while the quantity $W_{ij}$ is given by 
% \begin{equation}
    % W_{ij}=\displaystyle \frac{1}{2} \left(\frac{\partial \overline{u}_i}{\partial x_j}- \frac{\partial \overline{u}_j}{ \partial x_j} \right).
% \end{equation}
with $\kappa$ the von K\'arm\'an constant. The destruction term in (\ref{SAequation1}) captures the blocking effect of the wall on turbulent fluctuations and is a function of the distance to the nearest surface $d$. With this term, the model produces an accurate log-layer in wall-bounded flows. It includes
% The term in $\tilde{\nu}^2$ is the destruction term and it was built in order to produce an accurate log layer. 
a nondimensional function $f_{w}$  that increases the decay of the destruction term in the outer region. This term reads as
\begin{equation}
    f_{w} = g \left[ \displaystyle \frac{1+ c_{w3}^6}{ g^6 + c_{w3}^6}\right]^{1/6}
\end{equation}
with 
\begin{equation}
    g = r+c_{w2} \left( r^6 - r\right) \quad  \mathrm{and} \quad 
    r = \frac{\tilde{\nu}}{ \tilde{\mathcal{S}} k^2 d^2}.
\end{equation}
Standard values for the calibration constants $c_{v1}=7.1$ $c_{b1}=0.1355$, $\sigma=2/3$, $c_{b2}=0.622$, $c_{w2}=0.3$, $c_{w3}=2$ are used \citep{spalart1994}, with $c_{w1}=c_{b1}/\kappa^2+(1+c_{b2})/\sigma$ to balance production, diffusion and destruction in the log-layer,
 with $\kappa=0.41$.}
% It is worth noting that the balance between the transport terms in \eqref{SAequation1} depends quite strongly on the field variable $d$, the distance between a point in the computational domain and the nearest wall. As discussed later in section \ref{sec:modelling-surface-roughness}, the rough wall is treated by allowing the turbulent flow to develop in the wall-normal direction from a new virtual origin, shifted by $d_0$ beneath the numerical boundary. This shift alters the effective distance from the wall and needs to be taken into account in the SA model equations.

\subsection{Roughness model for homogeneous surfaces}\label{sec:roughness-model}
Many rough wall modelling strategies for homogeneous roughness for RANS simulations rely on the notion of equivalent sandgrain roughness $k_s^+$ (e.g., among others, \citet{Durbin2001, suga2006, Aupoix2007, BRERETON201874, prakash2020}). These strategies, described in this section, link the equivalent sandgrain roughness to suitable non-zero turbulence quantities (the modified eddy viscosity for the SA model) at the smooth, planar boundary of the numerical domain, to capture the increased turbulence activity near the rough surface and obtain the desired shift of the logarithmic {velocity} profile {as the main effect of the surface roughness}. No-slip boundary conditions are applied for the velocity. The mean turbulence structure is assumed to develop from a new virtual origin, displaced beneath the numerical boundary by a suitable distance $d_{0}^+$, to be determined \citep{rotta1962}. Relying on the outer-layer similarity hypothesis of \citet{townsend1976structure}, far away from the surface the law of the wall is preserved, and the shift of the streamwise velocity profile observed over a rough surface is captured by the empirical relation
\begin{equation}\label{eq:empirical-u}
    {\left.\frac{\partial \overline{u}_{1,r} }{\partial x_2^+}\right|_{x_2^+} = \left.\frac{\partial \overline{u}_{1, s}}{\partial x_2^+}\right|_{x_2^+ + d_0^+},}
\end{equation}
where the subscripts $(\cdot)_r$ and $(\cdot)_{s}$ denote quantities over the rough and smooth walls. Integrating this relation from the wall with {$\overline{u}_{1,r}(x_2^+=0) = 0$} yields 
\begin{equation}
    {\overline{u}_{1, r}(x_2^+) = \overline{u}_{1, s}(x_2^+ + d_0^+) - \overline{u}_{1, s}(d_0^+),}
\end{equation}
which evaluated far away from the surface gives the logarithmic shift
\begin{equation}\label{eq:du-virtual-origin}
    {\Delta \overline{u}_1 = \overline{u}_{1, r}(x_2^+ \gg d_0^+) - \overline{u}_{1, s}(x_2^+ \gg d_0^+) = \overline{u}_{1, s}(d_0^+)}
\end{equation}
i.e. the distance $d_0^+$ can be found as the wall-normal coordinate where the velocity over the smooth wall is equal to the desired velocity shift {$\Delta \overline{u}_1$.}

Then a roughness function that links the equivalent sandgrain roughness to the shift of the log region is required. Among various options available, here we use the Colebrook-Grigson roughness function \citep{grigson1992}, given by
\begin{equation}
    {\Delta \overline{u}_1= \frac{1}{\kappa} \log \left( 1+ \frac{k_s^+}{ \exp(3.25 \kappa)} \right),}
    \label{eq:colebrook}
\end{equation}
with $\kappa$ being the von K\'{a}rm\'{a}n {constant.} {Although significant variations can be observed in the transitional regime, little practical difference are found for the fully-rough regime in using this and other models, such as Nikuradze's roughness function \citep{Aupoix2007}. }
%Little practical difference in using this models and others, such as Nikuradze's roughness function \citep{Aupoix2007}, is observed in the fully-rough regime which we focus on in this paper, although significant variations can be observed in the transitional regime. 
Knowing the smooth-wall velocity profile $\overline{u}_{1, s}(x_2^+)$ and equating the relations \eqref{eq:colebrook} and \eqref{eq:du-virtual-origin} allows the displacement $d_0^+$ to be expressed as a function of the desired sandgrain roughness $k_s^+$. 

With such information, a solution consistent with \eqref{eq:empirical-u} can be found when the eddy viscosity satisfies
% \begin{equation}\label{eq:nu_bc_pre}
    % \nu_r(y^+) = \tilde{\nu}_s(y^+ + d_0^+).
% \end{equation}
% First, the inhomogeneous boundary condition
% \begin{equation}\label{eq:bc-nu}
%     \tilde{\nu}_r( y^+=0 ) = \tilde{\nu}_s(d_0^+) = d_0^+ \left. \frac{\partial \tilde{\nu}_s}{\partial y^+}\right|_{y^+=0}
% \end{equation}
% is enforced at the wall, where the second equality stems from the fact that the modified eddy viscosity in the SA model behaves linearly by construction from the wall through the log layer \citep{spalart1994}.
%----------------
% For instance, 
% i.e.
% \begin{equation}\label{eq:d0_from_hs}
%     d_0^+ = \left(1+\frac{k_s^+}{\exp(3.25 \kappa)}\right) \exp(-\kappa A),
% \end{equation}
% which is only valid in the fully rough regime.
%----------------
% First the eddy viscosity distribution is then assumed to be
\begin{equation}\label{eq:nu_bc_pre}
    \nu_{t, r}(x_2^+) = \nu_{t, s}(x_2^+ + d_0^+) 
    % = (y + d_0) \kappa,
\end{equation}
This is achieved in two steps. 
% where the subscripts $(\cdot)_r$ and $(\cdot)_{s}$ denote quantities for the rough and smooth wall cases and where $\kappa$ is the von K\'arm\'an constant. The second equality stems from the fact that the eddy viscosity distribution produced by the SA model will vary linearly by construction from the wall through the log layer, as the viscous sublayer has disappeared.
First, the inhomogeneous boundary condition
\begin{equation}\label{eq:bc-nu}
    \tilde{\nu}_r(x_2^+=0) = \tilde{\nu}_s(d_0^+) = d_0^+ \kappa / \Rey_\tau
\end{equation}
is enforced to the modified eddy viscosity in the SA model, where the second equality stems from the fact that, {in the SA model}, $\tilde{\nu}$ varies linearly as $\kappa x_2$ near the wall {by construction}. Second, the distance $d$ between any point in the computational domain and the nearest wall appearing in the SA model, as a fundamental field variable that controls the balance between the production and destruction terms, needs to be updated to reflect the location of the new virtual origin, slightly below the numerical domain boundary. {For instance, for the lower half of the channel, $d = x_2 + d_0$.} Overall, this procedure allows the SA model to produce the desired shifted logarithmic velocity profile, {consistent} with the updated boundary condition \eqref{eq:bc-nu} on the modified eddy viscosity.

\begin{figure}
    \centering
    \includegraphics[width=\textwidth]{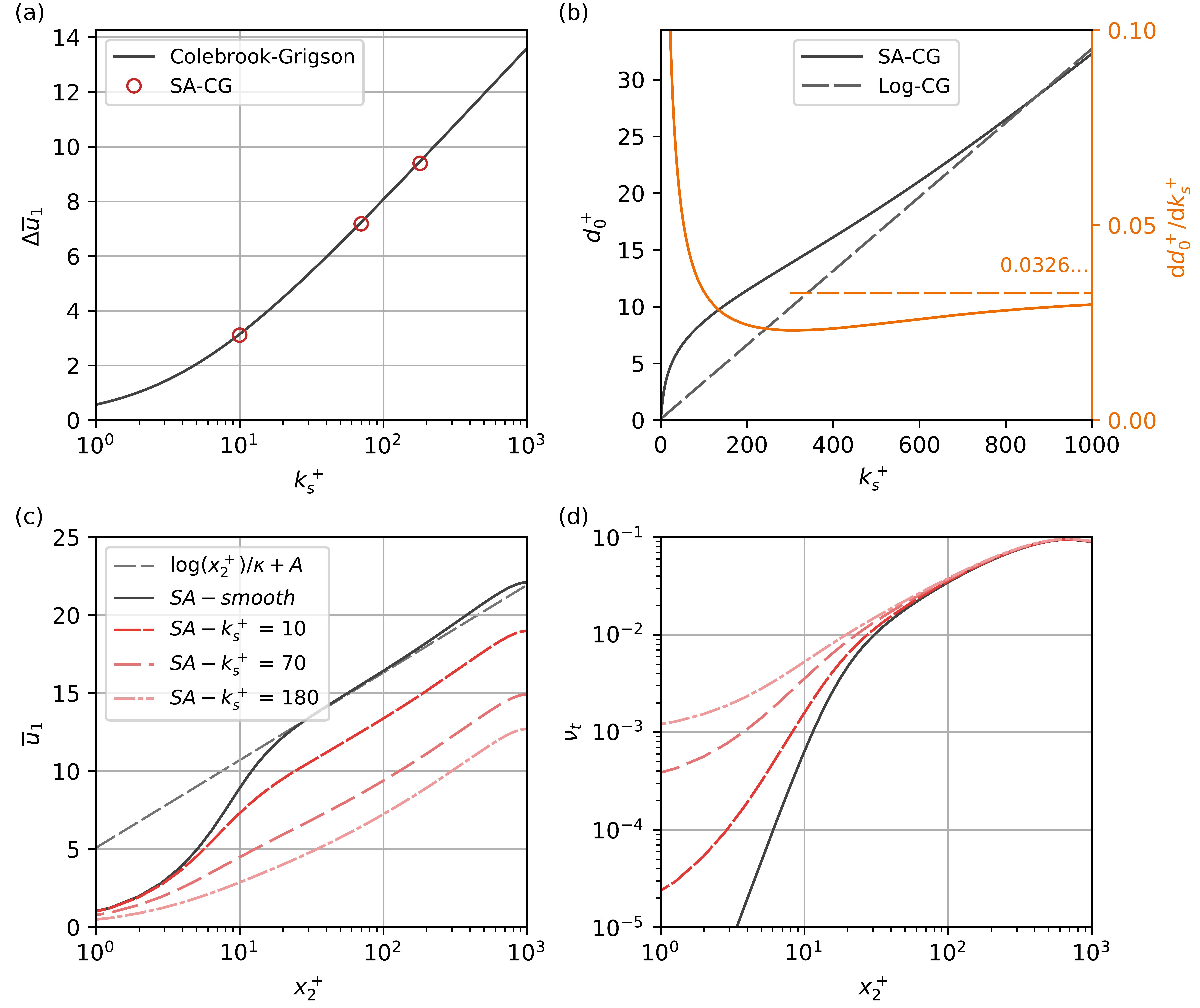}
    \caption{Panel (a), the analytical Colebook-Grigson roughness function (solid line) and the logarithmic shift obtained via the procedure described in the text (open circles). Panel (b), virtual origin $d_0^+$ as a function of the equivalent sandgrain roughness $k_s^+$ obtained by using the smooth-wall SA velocity profile or the standard log-law (black lines). The orange lines denote the derivative of $d_0^+$ with respect to $k_s^+$ (axes on the right hand side).
    {Numerical solutions of the streamwise momentum equation and the SA turbulence model} for channels with smooth and homogeneous rough walls for $\Rey_\tau = 1000$. Panel (c) streamwise velocity, panel (d), turbulent viscosity. The circles in panel (a) indicate the logarithmic shift associated to the profiles of panel (c), calculated using the procedure in the text.  For the logarithmic law, the constants $\kappa = 0.41$ and $A = 5.1$ are used.}
    \label{fig:zero_order_rough}
\end{figure}

Figure \ref{fig:zero_order_rough} illustrates example results at $\Rey_{\tau}=1000$, at which most of the results presented in later sections were obtained.
% The streamwise velocity and the turbulent viscosity at order zero, i.e. the solution over the homogeneous channel with equivalent sandgrain roughness $k_s^{(0)}$, appear as coefficients in the equations at order one. 
Solutions were obtained numerically with an in-house RANS code, based on a {Chebyshev}-collocation discretization method. {Mesh independence studies, omitted here, showed that 252 collocation points where sufficient to obtain mesh-independent results}. The nonlinear system of algebraic equations formed by the streamwise momentum equation and the SA equation was solved using a Jacobian-free Newton–Krylov technique \citep{knollkeyes2004}, using the ``hookstep'' approach of \citet{VISWANATH_2007} to improve convergence. Initial guesses for the streamwise velocity were obtained by first solving the momentum equation using Cess's analytical eddy viscosity profile {\citep{reynolds1972}}. For a desired equivalent sandgrain roughness $k_s^+$ for the homogeneous surface, the Colebrook-Grigson roughness function in panel (a) of figure \ref{fig:zero_order_rough} is first used to obtain $\Delta \overline{u}_1$. Using the smooth-wall velocity profile obtained from the SA model, panel (c), the virtual origin $d_0^+$ is obtained upon applying \eqref{eq:du-virtual-origin} {(panel b)}. Repeating this procedure for several equivalent sandgrain roughness yields the curve ``SA-CG'' in panel (b). Clearly, the choice of the smooth-wall velocity profile influences the results. For instance, coupling the Colebrook-Grigson formula {to} the log-law $\overline{u}_{1, s}(x_2^+) = \log (x_2^+)/\kappa + A$, with $\kappa = 0.41$ for consistency with the standard SA model and $A=5.1$, yields
\begin{equation}\label{eq:d0_from_hs}
    d_0^+ = \left(1+\frac{k_s^+}{\exp(3.25 \kappa)}\right) \exp(-\kappa A) \simeq 0.0326 k_s^+,
\end{equation}
denoted as ``Log-CG'' in the figure. 
% the difference \mbox{$u_s^+(y^+) - u_r^+(y^+)$} evaluated at some large distance from the wall, $y^+ \gg d_0^+$, then provides the shift of the logarithmic profile as a function of the displacement of the virtual origin:
% \begin{equation}\label{eq:du-virtual-origin}
%     \Delta u^+ = \frac{1}{\kappa} \log d_0^+ + A,
% \end{equation}
% i.e. the shift in the log region is equal to the smooth-wall velocity evaluated at a distance equal to the displacement of the virtual origin.
This virtual origin is then used for the boundary condition 
\eqref{eq:bc-nu} and in the SA model. Overall, this yields shifted velocity profiles, panel (c), that produce the desired $\Delta \overline{u}_1$ {as a function of $k_s^+$. This is demonstrated in panel (a), which compares the Colebrook-Grigson roughness function (solid line) used at the first step with the logarithmic shift obtained at the last step of this procedure for three values of $k_s^+$, denoted by the circles.}
% Calculations performed at $\Rey_{\tau}=1000$ are reported in figure \ref{fig:zero_order_rough}, showing streamwise velocity and turbulent viscosity profiles for a few roughness heights along with the smooth wall solution. 
It is worth pointing out that small {absolute} variations of the turbulent viscosity distribution in panel {(d)} are sufficient to produce relatively significant alterations of the mean velocity profile. Numerically, this makes the equations relatively stiff to solve.

\subsection{Roughness model for heterogeneous surfaces}\label{sec:modelling-surface-roughness}
% In the present work, the roughness geometry of the two strips is not directly resolved. Rather, we assume that the roughness elements on the two strips are much smaller than the channel height and the flow is then averaged over several roughness elements without resolving the flow details in their vicinity. 
% To model the roughness, we start by adopting the formalism of the equivalent sandgrain roughness $k_s$, normalised here with [FIX], introduced by \citet{nikuradse1933} as the diameter of semi-spheres distributed over a plane surface and producing the same shear stress and downward shift of the logarithmic profile \citep{jimenez2004, Flack2014}. 

The equivalent sandgrain roughness is a dynamic parameter that is non trivially related to the roughness geometry. For homogeneous roughness, it can be readily estimated from correlations once the shift of the velocity profile is known. However, for heterogeneous roughness, e.g. the present surface with alternating strips, it is not immediately clear how one should assign an equivalent sandgrain roughness to the two strips from velocity measurements, as the flow structure and thus the logarithmic shift also depend on the spatial distribution of the roughness properties \citep{wangsawijaya2020}. This conundrum is fundamentally the same discussed in numerical simulation studies in which the roughness is not resolved but suitable boundary conditions are applied at the smooth, planar boundary of the numerical domain. In such strategies, the roughness heterogeneity can be modelled directly by a lateral variation of the shear stress \citep{chung2018} or by a lateral variation of the transversal slip length \citep{Neuhauser_Schafer_Gatti_Frohnapfel_2022}. These strategies require a model that links the boundary conditions to the desired logarithmic shift. Such models are generally derived for homogeneous surfaces and their applicability to heterogeneous surfaces may be questioned.

\begin{figure}
    \centering
    \includegraphics[width=0.95\textwidth]{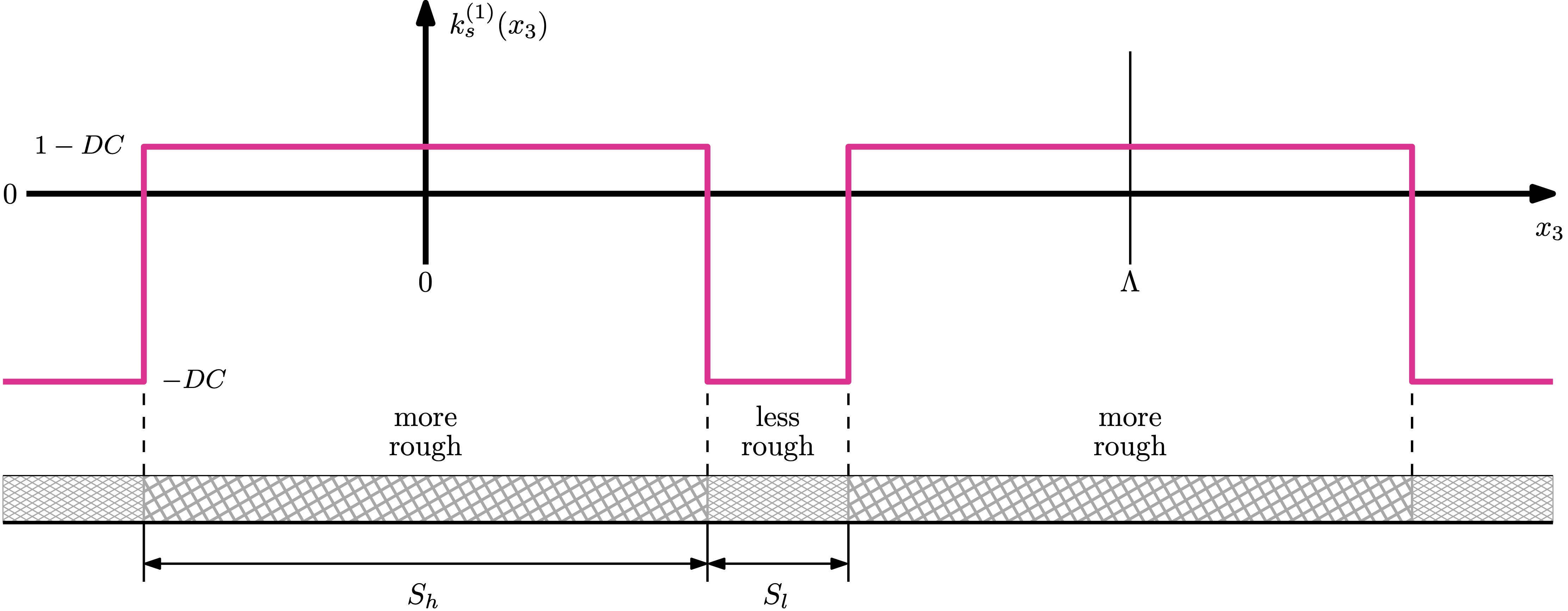}
    \caption{{The hashed diagram on the bottom shows the heterogeneous rough surface with roughness strips having higher or lower roughness than average, covering about two periods of the laterally repeating pattern. The duty cycle is $0.8$. The diagram at the top shows the distribution of the perturbation roughness height $k_s^{(1)}(x_3)$. Note that this term represents the deviation from the reference roughness height $k_s^{(0)}(x_3)$, as defined by equation \eqref{eq:hs}.}}
    \label{fig:order-one-roughness}
\end{figure}

Here, given the lack of a better strategy, the aforementioned approach is adopted. Specifically, the alternating strips are defined by a spanwise variation of the equivalent sandgrain roughness, following the expression
\begin{equation}
k_s^+(x_3) = k_s^{(0)}+\epsilon k_s^{(1)}(x_3),
\label{eq:hs} 
\end{equation}
where $k_s^{(0)}$ is the reference, spatially-constant roughness height and $k_s^{(1)}(x_3)$ captures the variation of the roughness properties over the two strips. For a {unitary {$k_s^{(1)}$} amplitude of the}
%one unit of the alternating 
roughness pattern, the spanwise variation is defined by the unitary peak-to-peak amplitude, zero-mean function
\begin{equation}\label{fx3}
    \displaystyle
    k_s^{(1)}(x_3) = \begin{cases}
                      1 - DC &  0 \leq x_3 \leq S_h/2 ~\mathrm{and} ~ \Lambda - S_h/2 \leq x_3 \leq \Lambda\\
                      - DC &  S_h/2 \leq x_3 \leq \Lambda - S_h/2,
                      \end{cases}
\end{equation}
{as demonstrated in the diagram of figure \ref{fig:order-one-roughness}}.
Because the difference in roughness between the two strips defined by $k_s^{(1)}(x_3)$ is unitary, the parameter $\epsilon$ in equation \eqref{eq:hs} controls the actual difference in roughness between the two strips, {although it is not related to the physical structure of the roughness.} This definition {is preferred over  specifying the roughness of the two strips, or considering roughness strips separated by smooth regions \citep{wangsawijaya2020}. This choice is} motivated by the fact that we consider in {the present analysis} the asymptotic limit when $\epsilon$ tends to zero, so that linearised equations governing the response of the turbulent shear flow developing over a rough surface to a small spanwise variation of the roughness properties can be obtained.

\subsection{Linearisation of the Reynolds-averaged equations and of the roughness model}
The streamwise momentum, streamfunction and SA equations form a coupled system of three nonlinear partial differential equations that can be solved for any desired strip configuration. However, when the difference between the properties of the high and low roughness strips is small, in the limit when $\epsilon \ll 1$, the resulting flow structure can be thought of as the sum of the flow in a channel with the homogeneous reference roughness $k_s^{(0)}$ and a small perturbation, produced by the surface roughness heterogeneity $k_s^{(1)}(x_3)$ and capturing the heterogeneous flow structure of secondary flows. This small perturbation obeys a set of linear equations which is much easier to solve, and only captures linear input-output mechanisms. To derive such equations, a generic time-averaged flow quantity $\overline{q}(x_2, x_3)$ is first expanded {in series} as
\begin{equation}\label{lineardecomposition}
    {\overline{q}(x_2, x_3) = q^{(0)}(x_2) + \epsilon {q}^{(1)}(x_2, x_3) + \mathcal{O}(\epsilon^2).}
\end{equation}
{Higher-order terms in \eqref{lineardecomposition} are neglected within the current framework. The convergence of the series for finite $\epsilon$ and the validity of the resulting predictions must still be verified. However, this approach is motivated by the goal of calculating the linear response of the turbulent mean flow to a small, nearly infinitesimal, perturbation in surface attributes, to assess how well linear mechanisms can account for the formation of secondary structures over heterogeneous surfaces.}

% The term $q^{(0)}(x_2)$ denotes the time-averaged solution in the plane channel with the reference homogeneous roughness. Considering the limit $\epsilon \ll 1$, when the difference between the properties of the strips is small, higher-order terms in $\epsilon$ in \eqref{lineardecomposition} are neglected and the term ${q}^{(1)}(x_2, x_3)$ captures the organisation of secondary currents induced by the heterogeneous surface, 
% \emph{for a unitary difference in equivalent sandgrain roughness between the two strips}. 
% 
Substituting this expression for all mean quantities in the Reynolds-averaged equations and in the SA equation, and taking terms at order zero in $\epsilon$, leads to the nonlinear equations governing the flow over the homogeneous rough surface. 
{The streamwise vorticity equation is trivially satisfied by $\psi^{(0)} = 0$. The streamwise momentum equation is
\begin{equation}
\displaystyle
0 = 1 + \displaystyle\frac{1}{\Rey_\tau} \frac{\partial^2 u_1^{(0)}}{\partial x_2^2} + \frac{\partial \tau_{12}^{(0)}}{\partial x_2}, 
\label{systemzeroorder}
\end{equation}
and it is coupled to the SA transport equation via the definition of the Reynolds stress $\tau^{(0)}_{12}$. These two equations are solved in a coupled fashion} using the approach discussed in section \ref{sec:roughness-model}. At first order, the equations governing the perturbation of the streamwise velocity and the streamfunction are
\begin{subeqnarray}\label{eq:linear_equations}
\displaystyle
\hspace{-0.8cm}-\frac{\partial \psi^{(1)}}{\partial x_3} \Gamma &&\!=\!\frac{1}{\Rey_\tau}\!\left(\frac{\partial^2 }{\partial x_2^2}\!+\!\frac{\partial^2 }{\partial x_3^2}\right) u_1^{(1)}\!+\!\frac{\partial \tau_{12}^{(1)} }{\partial x_2}\!+\!\frac{\partial \tau_{13}^{(1)} }{\partial x_3},
    \label{OSfinal}\\
    \displaystyle
  0&&\!=\!\frac{1}{\Rey_{\tau}}\!\left( \frac{\partial^2 }{\partial x_2^2}\!+\!\frac{\partial^2}{\partial x_3^2}\right)^2\! \psi^{(1)}\!+\!\frac{\partial^2}{\partial x_2 \partial x_3} \left( \tau_{33}^{(1)}\!-\!\tau_{22}^{(1)}\right)\!+\!\left( \frac{\partial^2}{\partial x_2^2}\!-\!\frac{\partial^2}{\partial x_3^2}\right)\! \tau_{23}^{(1)},
\label{psiequation}
\end{subeqnarray}
where we define $\Gamma=\partial u_1^{(0)}/\partial x_2$, showing that the zero-order solution, through the mean velocity gradient $\Gamma$, needs to be available for the solution of the first-order equations. {The first term at the left-hand side of (\ref{OSfinal}a), analogous to the off-diagonal coupling operator in the Orr-Sommerfeld-Squire linearised equations, is the only coupling term explicitly appearing in this set of equations. Physically, this terms captures the interaction between the mean shear and the perturbation velocity and underpins energy extraction mechanisms in shear flows via the lift-up effect \citep{brandt2014}.} {All other terms obtained from the nonlinearity vanish because the streamfunction at order zero is identically zero.} {Secondary currents also introduce an alteration of the spatial organisation of the turbulent viscosity through an alteration of the balance of the transport terms in the SA equation. The linearised SA equation governing such organisation is coupled to the streamwise momentum and the streamfunction equations and contributes to the perturbation of the Reynolds stresses entering \eqref{eq:linear_equations}.} {Linearisation of the SA model is tedious and leads to complex expressions. More detail on the linearisation procedure is reported in our previous work (see appendix B of \citet{zampino2022}), and is omitted here for brevity.}

% Considering the first-order terms, the 

It is worth noting that the streamfunction equation contains the perturbation of the Reynolds stresses originating from the cross-stream velocity components, as is well known \citep{perkins1970}. {Although at} order zero these {terms exhibit negligible influence}, at order one the perturbation of the Reynolds stress tensor {becomes pivotal to couple the two equations in the differential system \eqref{eq:linear_equations}. Here, the first-order stresses are} found by expanding the nonlinear Reynolds stress model \eqref{eq:taunonlinear} in a Taylor series in $\epsilon$, {leading to
\begin{equation}\label{firstordertau}
%   \hspace{-0.8cm}\tau_{ij}^{(0)}&\!=\! &  2\nu_t^{(0)}S_{ij}^{(0)}\!-\!c_{r1}\!\left[O_{ik}^{(0)}\tau_{jk}^{L(0)}\!+\!O_{jk}^{(0)}\tau_{ik}^{L(0)}\right],\\
%   	\tau_{ij}^{(1)} & \!= \!& 2\nu_t^{(0)}S_{ij}^{(1)}\!+\!2\nu_t^{(1)}S_{ij}^{(0)} 
%   	 \!-\!c_{r1}\!\left[O_{ik}^{(1)}\tau_{jk}^{L(0)}\!+\!O_{ik}^{(0)}\tau_{jk}^{L(1)}\!+\!O_{jk}^{(1)}\tau_{ik}^{L(0)}\!+\!O_{jk}^{(0)}\tau_{ik}^{L(1)}\right].
   	%  \hspace{-0.8cm}\tau_{ij}^{(0)}&\!=\! &  \tau_{ij}^{L(0)}\!-\!c_{r1}\!\left[O_{ik}^{(0)}\tau_{jk}^{L(0)}\!+\!O_{jk}^{(0)}\tau_{ik}^{L(0)}\right],\\
   	\tau_{ij}^{(1)} =  \tau_{ij}^{L(1)}
   	 - c_{r1}\!\left[O_{ik}^{(1)}\tau_{jk}^{L(0)} + O_{ik}^{(0)}\tau_{jk}^{L(1)} + O_{jk}^{(1)}\tau_{ik}^{L(0)} + O_{jk}^{(0)}\tau_{ik}^{L(1)}\right],
\end{equation}
where $O_{ij}^{(1)}$ is the normalised rotation tensor induced by the first-order velocity components (see appendix \ref{appendix:qcr}). 
% Note that the tensor $\tau_{ij}^{(1)}$ is still symmetric although the tensor $O_{ij}^{(1)}$ is antisymmetric.
Developing (\ref{firstordertau}), the individual perturbation Reynolds stresses appearing in \eqref{eq:linear_equations} are
\begin{subeqnarray}
\label{tau23qcr}
    	\tau_{12}^{(1)}&=&\nu_t^{(0)}\displaystyle \frac{\partial u_1^{(1)}}{\partial x_2}+ \nu_t^{(1)}\Gamma+2c_{r1} \mathrm{sign}(\Gamma)\nu_t^{(0)} \displaystyle \frac{\partial^2 \psi^{(1)}}{\partial x_2 \partial x_3},\\
	\tau_{13}^{(1)}&=&\nu_t^{(0)} \displaystyle \frac{\partial u_1^{(1)}}{\partial x_3}-2c_{r1} \mathrm{sign}(\Gamma)\nu_t^{(0)} \frac{\partial^2 \psi^{(1)}}{\partial x_2^2},\\
	\tau_{23}^{(1)}&=&\nu_t^{(0)}\left(\displaystyle \frac{\partial^2}{\partial x_2^2}- \frac{\partial^2}{\partial x_3^2}\right) \psi^{(1)}+2 c_{r1} \mathrm{sign}(\Gamma) \nu_t^{(0)} \frac{\partial u_{1}^{(1)}}{\partial x_3},\\
	\tau_{22}^{(1)}&=&-2 \nu_t^{(0)} \displaystyle \frac{\partial^2 \psi^{(1)}}{\partial x_2 \partial x_3}+2 c_{r1} \left[ \mathrm{sign}(\Gamma)\nu_t^{(0)} \displaystyle \frac{\partial u_{1}^{(1)}}{\partial x_2}+\mathrm{sign}(\Gamma) \nu_t^{(1)}\Gamma\right],\\
	\tau_{33}^{(1)}&=&2 \nu_t^{(0)} \displaystyle \frac{\partial^2 \psi^{(1)}}{\partial x_2 \partial x_3}.
\end{subeqnarray}
Except for $\tau_{33}^{(1)}$, which coincides with its linear Boussinesq's definition, all other stresses contain an additional term specific to the QCR model and proportional to the $c_{r1}$ constant. In particular, the stresses appearing in the streamfunction equation contain spatial gradients of the streamwise velocity, and vice versa. These terms result in a tighter, two-way coupling between the streamfunction and streamwise velocity equations, now able to sustain secondary currents.}

% {After the linearisation,} these stresses depend on both the perturbation streamwise velocity and the perturbation streamfunction, {leading to} 
% %strongly coupling the two equations in \eqref{eq:linear_equations}. Overall, this leads to 
% a set of three coupled linear differential equations.

% Overall, a set of three linear differential equations is obtained

% To account for the new equilibrium, the modified viscosity is expanded similarly to the other flow variables as
% \begin{equation}
%     \tilde{\nu}(x_2, x_3) = \tilde{\nu}^{(0)}(x_2) + \epsilon \tilde{\nu}^{(1)}(x_2, x_3) + \mathcal{O}(\epsilon^2),
%     \label{eq:nu_t_definition}
% \end{equation}
% where the first-order term $\tilde{\nu}^{(1)}$ now captures the altered spatial distribution of the turbulence viscosity, compared to the homogeneous roughness distribution given by the zero-order term $\tilde{\nu}^{(0)}$. 

To obtain boundary conditions for the field variables, the wall roughness treatment model discussed in section \ref{sec:modelling-surface-roughness} needs to be linearised. The key idea is that small spanwise perturbations of the equivalent sandgrain roughness are modelled as small spanwise variations of the virtual origin. More formally, over the heterogeneous surface given by \eqref{eq:hs}, the shift of the virtual origin varies according to
\begin{equation}
    d_{0}^+(x_3) = d_{0}^{(0)} + \epsilon d_{0}^{(1)}(x_3)
\end{equation}
where $d_0^{(0)}$ is the shift of the virtual origin of the reference homogeneous surface with equivalent sandgrain roughness $k_s^{(0)}$. On the other hand, the first-order term can be found by differentiating numerically the curve reported in figure {\ref{fig:zero_order_rough}(b)} at $k_s^+= k_s^{(0)}$ and using \eqref{fx3}, since
\begin{equation}\label{eq:linearised-d-h-relation}
    d_0^{(1)} = \left. \frac{\mathrm{d} d_0^+}{\mathrm{d} k^+_s}\right|_{k_s^{(0)}} k_s^{(1)}.
\end{equation}
Asymptotically, for large $k_s^{(0)}$ and considering the log law, the derivative of the curve in figure {\ref{fig:zero_order_rough}(b)} tends to about 0.0326, implying that a peak-to-peak variation of the equivalent sandgrain roughness of $1/0.0326 \approx 30$ is necessary to obtain a peak-to-peak variation of the virtual origin equal to the viscous length scale. Once $d_{0}^{(1)}$ is known,  {linearising \eqref{eq:bc-nu} yields the wall condition for the modified eddy viscosity}
\begin{equation}
    \tilde{\nu}^{(1)}(x_2 = 0) = d_0^{(1)} \kappa / \Rey_\tau,
    \label{eq:BCs_first}
\end{equation}
showing that the spanwise variation of the roughness properties is modelled as a lateral change in the eddy viscosity at boundary of the numerical domain. Finally, homogeneous boundary conditions are used for the streamwise velocity perturbation and the streamfunction perturbation and its wall-normal derivative. 

One last remark is in order. A sensible question is whether the inclusion of the linearised turbulence model to describe the perturbation of the turbulent viscosity is really necessary, given that this is not customary in many previous studies using linearised Navier-Stokes equations. On the one hand, the strength of the mean flow response to a lateral perturbation of the surface attributes {may likely depend on the well-known selective amplification properties of the linearised Navier-Stokes operator, which are largest when such perturbation occurs at a specific spanwise length scale}. On the other hand, the SA model provides a means to model realistically the effect of the surface heterogeneity, because it provides clear insight into how the perturbation of the effective distance $d$ influences the perturbation of the turbulent viscosity field. As described in section \ref{sec:combination}, the lateral perturbation of the distance $d$ is the dominant source mechanism that leads to secondary flows in the present framework. In principle, one could first introduce an ansatz on the perturbation of the turbulent viscosity and then only solve the linearised Navier-Stokes equations. However, given the range of transport phenomena modelled by the SA equation, defining the correct ansatz does not appear to be a straightforward task.

\subsection{Numerical solution of the linearised equation} \label{sec:fourier}
The spanwise variation of the equivalent roughness height can be modeled as a square wave approximated by the cosine series
\begin{equation}\label{eq:fx3_approximation}
    k_s^{(1)}= \sum_{n=1}^{\infty} \, k_s^n \cos \left (n \frac{2\pi}{\Lambda} x_3 \right).
\end{equation}
The coefficients $k_s^n$ can be calculated analytically for each combination of widths $S_h$ and $S_l$ and the corresponding coefficients $d_0^n$ for the spanwise variation of the virtual origin $d_0^{(1)}$ are found by using \eqref{eq:linearised-d-h-relation}. 

Expanding the unknown field variables at first-order in series, e.g.~for the streamwise velocity
\begin{equation}\label{eq:expansion}
    u_1^{(1)} (x_2, x_3)  = \sum_{n=1}^{\infty} \, \hat{u}_1(x_2; n) \cos \left (n \frac{2\pi}{\Lambda} x_3 \right),
    % \\
    % \psi^{(1)} (x_2, x_3) &=&\sum_{n=1}^{\infty} \, \hat{\psi}(x_2; n) \sin{(n k_3 x_3)},\\
       % \tilde{\nu}^{(1)} (x_2, x_3) &=&\sum_{n=1}^{\infty} \, \hat{\nu}(x_2; n) \cos{(n k_3 x_3)}
\end{equation}
and % where $\hat{u}_1(x_2;n)$, $\hat{\psi}(x_2;n)$ and $\hat{\nu}(x_2;n)$ are the real-valued, wall-normal profiles of the perturbation streamwise velocity, streamfunction and modified eddy viscosity at each integer multiple of the fundamental wavenumber $k_3$. 
substituting these expressions in the linearised equations leads to one set of three linear ordinary differential equations in $x_2$ for each integer wavenumber. {As opposed to previous studies considering the linearised Navier-Stokes equations \citep{Chavarin.2020, Ran.2020}, each set of three ordinary different equations is independent of all other wavenumber and can be solved in isolation.} {This would not be the case if higher order terms had been retained in \eqref{lineardecomposition}, and a larger problem would need to be solved taking into account harmonic interactions.} {A Chebyshev-collocation method was used for the discretisation}. Although the field variable $d$ in the SA model has a sharp cusp at $x_2=0$ and hence a spectral technique is not ideal for the solution of this problem, we have observed that the numerical method is robust enough to provide accurate results when a sufficiently fine grid is used. In the following simulations, we used up to 252 collocation points. 
For the spanwise discretisation, we observed that solutions converge relatively rapidly with the number of Fourier modes retained in the expansion \eqref{eq:fx3_approximation}. This can be motivated by the observation that, far away from the wall, only large-scale perturbations of the surface features can influence the flow structure, while the effect of small-scale perturbations, i.e. sharp gradients of the boundary conditions \citep{Neuhauser_Schafer_Gatti_Frohnapfel_2022}, decays more rapidly with the distance from the wall \citep{meyers2019}. {The number of spanwise modes required increases with the fundamental wavelength $\Lambda$. We always checked that results did not change visibly when doubling the number of modes. As a reference, twenty modes were sufficient at $\Lambda \approx 1$ to obtain a converged description of the perturbation velocity field.}
% Homogeneous boundary conditions for the all velocity components are applied, while from   \eqref{eq:BCs_first}, the linearised boundary conditions for the modified eddy viscosity becomes
% \begin{equation}
%     \hat{\nu} (x_2=\pm 1; n) = d_0^n \kappa.
% \end{equation}

The final solution is then found by combining the solutions at each wavenumber, as the superposition principle applies. One important implication of this property is that the flow structure over surfaces with complex topographic/roughness characteristics \citep{mejia2013, barros2014} may be rationalised and better understood by decomposing the surface forcing into its constitutive components. The strength of the mean flow response at each spanwise length scale will then depend on the amplitude of such components times a factor that captures the selective amplification properties of the linearised Navier-Stokes operator \citep{chernyshenko_baig_2005}, augmented by the linearised SA equation.

% For more details see \citet{zampino2022}.
% Due to the linearity of the RANS-based model \eqref{eq:linear_equations}, the total field is hence given by  by the superposition of the results for a single wavelength mode.
%From \cite{zampino2022}, any topological heterogeneity produces secondary flows because of two separate forcing terms already discussed for the topographical heterogeneity. The first mechanism is the distance perturbation introduced by the virtual origin that affects the linear transport equation of the eddy viscosity and alter the source term in the near wall region where the transporting phenomena are stronger. In addition, the spanwise modulation of the eddy viscosity is recognised to be a source term in the turbulent kinetic energy density \citep{barros2014,Hwanglee2018}. The second forcing mechanism is given by the non-zero boundary condition for the eddy viscosity that account for the roughness effect. 
%This forcing is proportional to the coefficient $f^n$ that is a property of the surface strip configuration. 

It is worth noting that the spatially-constant component at $n=0$ does not appear in \eqref{eq:fx3_approximation} because of the assumption that the spanwise variation of the roughness height given by \eqref{fx3} is zero-mean. {It does also not appear in the solution \eqref{eq:expansion}, because the linearity of the model implies that a perturbation of the surface properties at wavenumber $n$ only produces a distortion of the time averaged flow at the same wavenumber.} A corollary of this property is that the present model does not predict any change in mean friction drag, the subject of several recent studies \citep{HUTCHINS2023113454, Frohnapfel_von_Deyn_Yang_Neuhauser_Stroh_Gatti_2024}. In fact, the spanwise-constant component $\hat{u}_1(x_2; 0)$ and thus the perturbation of the bulk velocity computed from this profile, {which would allow calculating the change in friction coefficient at constant friction velocity}, is identically zero. The model does indeed capture the spanwise modulation of the streamwise velocity distribution, i.e. high and low momentum pathways \citep{barros2014}, but second order effects in $\epsilon$ {that produce interactions between harmonics are necessary to obtain a velocity perturbation at wavenumber $n=0$ from surface perturbations at $n>1$, and thus capture the change in friction \citep{ZampinoPhD}}.

% Finally, each set of ordinary differential equations is solver numerically using a Chebyshev-collocation method. Although the field variable $d$ in the SA model has a sharp cusp at $x_2=0$ and hence a spectral technique is not ideal for the solution of this problem, we have observed that the numerical method is robust enough to provide accurate results when a sufficiently fine grid is used. In the following simulations, we used up to 252 collocation points. For the spanwise discretisation, we observed that solutions converge relatively rapidly with the number of Fourier modes retained in the expansion \ref{eq:fx3_approximation}. This can be motivated by the observation that, far away from the wall, only such large-scale perturbations of the surface features can influence the flow structure, while the effect of small-scale perturbations, i.e. sharp gradients of the boundary conditions \citep{Neuhauser_Schafer_Gatti_Frohnapfel_2022}, decays more rapidly with the distance from the wall \citep{meyers2019}. In practice, we checked that results did not change significantly when doubling the number of spanwise modes.

\section{Structure and strength of secondary currents} \label{sec:strips}
%[GIRARE LA FRASE COSì DA DIRE QUELLO CHE FAI E LA PROCEDURA. IL MODELLO CI PERMETTE DI FARE COSA? PER OGNI VALORE DI COPPIA WH ED Wl]

\begin{figure}
    \centering
    \includegraphics[width=\textwidth]{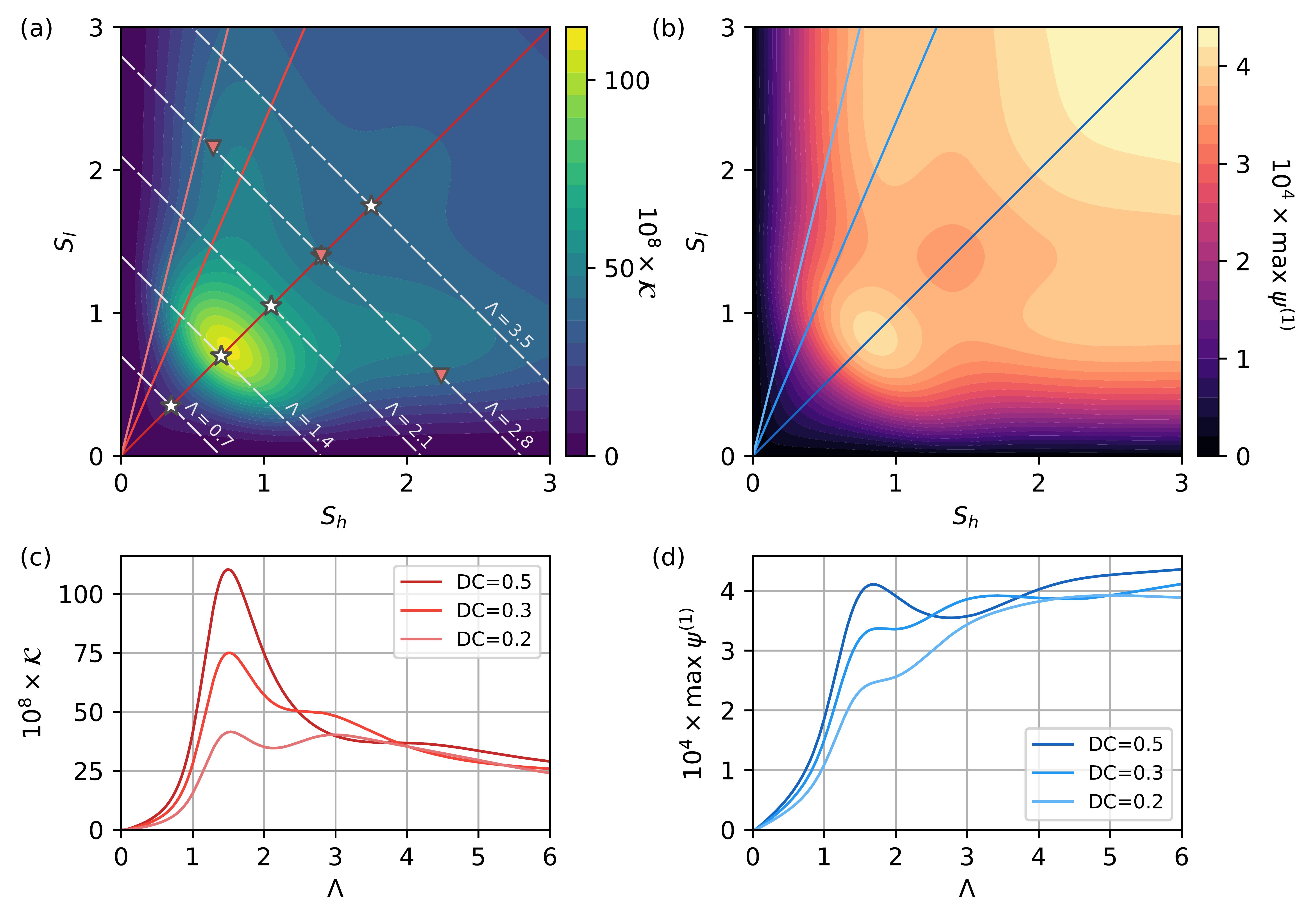}
    \caption{Contours of the volume averaged kinetic energy of the cross-stream velocities $\mathcal{K}$ (panel a) and the streamfunction peak value $\max_{x_2,x_3}|\psi^{(1)}|$ (panel b) as a function of the width of the high- and low-roughness strips. The Reynolds number is $\Rey_\tau=1000$ and $k_s^{(0)} = 180$. Cases at constant spatial fundamental length scale $\Lambda$ are identified by the dashed diagonal lines with negative slope. Markers identify cases discussed later in the text. Panels (c-d) shows the same two quantities for $DC=0.2, 0.3$ and 0.5, as a function of $\Lambda$.}
    \label{fig:maps}
\end{figure}

The volume averaged kinetic energy of the cross-sectional velocity components $\mathcal{K}$, defined as
\begin{equation}
\mathcal{K} = \frac{1}{4\Lambda}\! \int_{0}^{2} \int_{0}^{\Lambda} \left[ u_2^{(1)}(x_2, x_3)^2 + u_3^{(1)}(x_2, x_3)^2\right ] \,\mathrm{d}x_3 \,\mathrm{d}x_2,
\label{eq:kappa}
\end{equation}
is used here to characterise the strength of the secondary flows. We also use the streamfunction peak $\max_{x_2,x_3} |\psi^{(1)}(x_2,x_3)|$ to quantify the cross-stream flow rate associated with the vortices, as in other studies \citep{vidal2018}. Note that these variables are scaled with $u_\tau$ and $h$. The solution of the linearised equations for a given strip configuration can be obtained quite rapidly, which enables a rapid exploration of the parameter space $(S_h, S_l)$. Results are reported in the two top panels of figure \ref{fig:maps} for $\Rey_{\tau}=1000$, and using {$k_s^{(0)} = 180$}. Panels (c-d) show cuts along lines for three duty cycles as a function of the fundamental length scale $\Lambda$.
% \begin{figure}
%     \centering
%     \includegraphics[width=\textwidth]{maps_topology.eps}
%     \caption{Contours of the volume averaged kinetic energy of the cross-stream plane velocities $\mathcal{K}$ (panel a) and the streamfunction peak value $\max_{x_2,x_3}|\psi^{(1)}|$ (panel b) as a function of the width of the high- and low-roughness strips, indicated with $W_h$ and $W_l$ respectively. The Reynolds number is $\Rey_\tau=1000$. The cases at constant spacing $S$ are identified by the white lines. Dashed lines identify cases at constant gap or width, with markers for configurations discussed later in the text. [FIX: non capisco perche le linee tratteggiate sono cosi corte. Si possono toglere. e i simboli farli di un altro colore che si confonde con la mappa sotto.]}
%     \label{fig:maps}
% \end{figure}

We note that, as in our previous application of these techniques to surfaces with longitudinal ridges \citep{zampino2022}, the results of the linearised model become asymptotically Reynolds number independent for high Reynolds numbers, somewhat supporting the weak Reynolds number dependence documented in the literature \citep{wangsawijaya2022}. {This ultimately stems from known properties of the SA model \citep{spalart1994}, which is designed to produce an eddy viscosity distribution consistent with the log law}. Hence, the {discussion presented here can also be applied to} higher Reynolds number flows relevant to applications. Note also that flow variables, such as the velocity or streamfunction perturbations, are computed in the present linear modelling framework per unit variation of the equivalent sandgrain roughness height $k_s^{(1)}$ (scaled in inner units) in analogy to what was described for ridge-type roughness in \citet{zampino2022} where the same quantities are obtained per unitary ridge height (scaled in outer units). Given that experiments on secondary flows over heterogeneous surfaces are often conducted on roughness strips with a considerable difference in roughness properties \citep{wangsawijaya2020, chung2018}, the numerical values reported here will appear quite small. 
For graphical convenience, quantities are pre-multiplied by a large factor, e.g.~$10^8$ in \ref{fig:maps}(a), in the figures.

Regardless of the quantity used for measuring the strength of the secondary currents, a peak is observed for $S_h = S_l \simeq 0.7$, corresponding to a fundamental length scale $\Lambda \approx 1.4$, although the streamfunction peaks slightly later. Further, the quantities in figure \ref{fig:maps} are symmetric with respect to the line $DC=0.5$, where the strength peaks. For $\Lambda \gtrsim 2.5$ two peaks are observed, located symmetrically with respect to the line $DC=0.5$. Examination of the flow structure for some of these cases indicates that the secondary flows observed over the high- and low-roughness regions for a generic configuration $(S_h, S_l)$ are identical in strength but opposite in flow direction when the width of the two strips is swapped.  The strip width at which secondary currents are most intense reflects previous observations. For instance, \citet{chung2018}, using LES simulations, and \citet{wangsawijaya2020}, using experiments above spanwise-alternating smooth and rough strips, reported a maximum intensity for the secondary flows when the width of the strips is comparable with the boundary layer thickness. In particular, \citet{wangsawijaya2020} found that the swirl strength was largest among the cases they considered for $S_h = S_l = 0.62$. 

The contours of the kinetic energy around the peak region appear elongated along the line $S_h+S_l=const$. As a result, the three cuts along lines at constant duty cycle all display a peak for $\Lambda\simeq 1.4$, which may interpreted as $\Lambda$ being the relevant length scale. This is partly correct, as the response to sinusoidal perturbations of the sandgrain roughness does indeed peak for this length scale, as for ridge-type roughness \citep{zampino2022}. However, there is a marked effect of the relative size of the strips away from the peak and the two {widths} are indeed necessary to correctly characterise the response. The maps of figure \ref{fig:maps} show strong similarities with the maps displayed in \citet{zampino2022} {(see figure 10)} and \citet{Zampino2023} {(see figure 2)} for the ridge-type roughness, as the peak amplification occurs for similar values of $\Lambda$.
%In their paper, \citet{zampino2022} argued that the secondary flows generated by streamwise-aligned ridges occupy available space above/between the ridges until to their maximum size equal to the channel half-height. The maximum size is reached at $S=1.34$, corresponding to the amplification peak, but it slightly depends on the ridge shape. 
This similarity suggests that the selective amplification of secondary flows is an intrinsic property of the mean flow, and perhaps less strongly an effect of the type of forcing, e.g. whether it is produced by elevation (ridges) or roughness (strips) variations. In this regard, there has been recent discussion on the relation and co-existence between secondary currents and very large scale motions (VLSM, e.g. \citet{Lee_Sung_Adrian_2019}). One speculation is that secondary currents are naturally-occurring VLSMs that are phase locked spatially by the heterogeneous surface \citep{chung2018, wangsawijaya2020} and emerge in the time-averaged flow. The present approach, based on the Reynolds-averaged equations where the concept of VLSMs does not apply immediately, suggests that secondary currents may be interpreted as the time-averaged response of a forcing localised near the wall and produced by gradients of the turbulent stresses. The linearised Navier-Stokes operator with its selective amplification properties then produces more intense time-averaged structures at specific forcings wavelengths, when $\Lambda \simeq 1.4$. This hypothesis leverages the same physics used in transient growth analysis studies \citep{delalamo2006, cossu2009, pujals2009} to explain the formation of coherent structures in shear flows from properties of the Orr-Sommerfeld-Squire equations, except that here we consider the steady response to a given steady perturbation localised at the wall rather than the transient amplification of an optimal initial perturbation. 
% A further implication is that for real-world surfaces where multiple topographic length scales co-exist \citep{bons.2011, mejia2010}, e.g. with complex, broadband roughness patterns \citep{barros2014} that force the mean flow over a broad range of scales, secondary currents with size $\simeq 1.4$ the outer scale will dominate.

\subsection{Flow structure}
\begin{figure}
    \centering
    % \includegraphics[width=0.9\textwidth]{strips_topology_comparison_ver2.eps}
    %Davide se vuoi rifaccio anche la figura 10 per alcuni casi con DC=0.5 ed alcuni S
    \includegraphics[width=\textwidth]{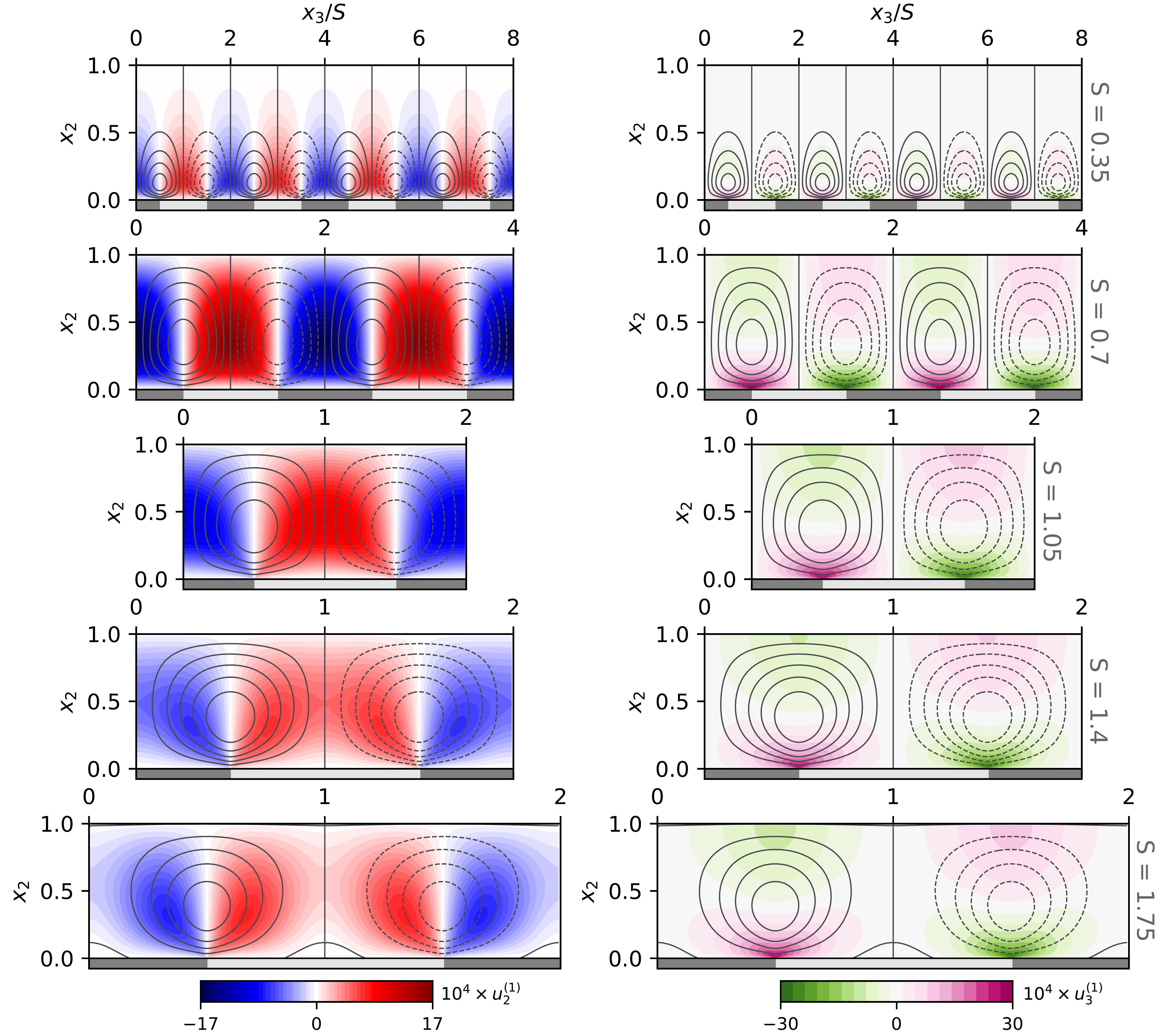}
    \caption{Colour maps of the wall-normal velocity perturbation $u_2^{(1)}$ over roughness strips at $DC=0.5$ for several strip widths $S$ (left column). The field in the fundamental domain (see figure \ref{fig:sketch}) is repeated four and two times for the two narrowest strip cases. Contour lines of the perturbation of the streamfunction $\psi^{(1)}$ are also reported to better describe the secondary flows. Negative $\psi^{(1)}$ are indicated by using dashed lines. Contours of the spanwise velocity perturbation $u_3^{(1)}$ (right column).
    The Reynolds number is $\Rey_{\tau}=1000$, for {$k_s^{(0)} = 180$}. The darker/lighter patches denote the high/low roughness strips.}
    \label{fig:topology}
\end{figure}

To visualise the cross-stream structure of the secondary currents, fields of the wall-normal and spanwise velocity perturbation are reported in figure \ref{fig:topology}, for duty cycle $DC=0.5$, i.e. for equal length of the high and low roughness strips, for a range of strip widths $S$. 
These configurations correspond to the star markers in figure \ref{fig:maps}(a). Contour lines of the streamfunction are also reported. The roughness strips produce two counter-rotating vortices inducing a down-welling over the high-roughness regions and an up-welling over the low-roughness regions. For narrow strips, even narrower than what shown here, the vortices are confined in the near wall region and the flow appears homogeneous at distances from the wall larger than the strip width. 
% From a physical standpoint, this can be explained by the amplifying properties of the linearised Navier-Stokes-SA system, where disturbances localised at the wall at a given spanwise length scale $S$ penetrate across the shear layer up to a depth $S$ into the wall-bounded flow. 
Increasing the strip width, the vortices grow with $S$ until they occupy the full half-height of the channel. For $S \simeq 0.7$, the cross-stream components and in particular the wall-normal component is more intense than all other cases considered. This configuration identifies the peak amplification region of the maps of figure \ref{fig:maps}. Increasing $S$ further, the strength of the wall-normal motions decreases slightly, but the volume averaged strength of such motions decreases much further as the flow structure converges to an idealised wide-strip asymptotic limit where moderately intense cross-stream motions are only found in the immediate vicinity of the transition between strips, with fluid is at rest in the ``homogeneous'' regions above the centres of the roughness strips. This explains the trend of the two quantities in figures \ref{fig:maps}{(c-d)}. While $\mathcal{K}$ is a volume averaged quantity and decreases with $S$ for $S\gg 1$, the streamfunction peak is a local quantity that measures the strength of the individual vortex cores. This quantity shows a first peak for $S \simeq 0.75$, and eventually tends to an asymptotic value (with a larger amplitude) for large $S$, characterising the strength of the ``isolated vortex'' regime.

% \begin{figure}
%     \centering
%     % \includegraphics[width=1.05\textwidth]{cross_section_velocity_v2.eps}
%     \includegraphics[width=1.05\textwidth]{profiles-u2-u3.jpg}
%     \caption{Profiles of the wall-normal (panel a) and spanwise velocity perturbations (panel b) for $h_s^{(0)} = $ [FIX] and unitary roughness height perturbation $h_s^{(1)}$ [FIX: ma hs plus quanto vale? ]. The velocity profiles are obtained at the centre of the high-roughness strips for $W_l=0.67$ and varying $W_h$ from 0.25 to 2, as reported in the legend of the figure. The Reynolds number is $\Rey_{\tau}=1000.$}
%     \label{fig:profile}
% \end{figure}

Near the transition between strips, the spanwise component is particularly intense in the near-wall region (right panels in figure \ref{fig:topology}).
% This is better seen in the wall-normal profiles of the two cross-stream components extracted at the centre of the high-roughness strips \mbox{($x_3=0$)}, reported in figure \ref{fig:profile} for the same strip configurations. 
The spanwise velocity peak is more intense than the vertical velocity peak in agreement with observations \citet{Frohnapfel_von_Deyn_Yang_Neuhauser_Stroh_Gatti_2024}, and is localised in the near wall region. Given that the spanwise velocity obeys no-slip condition, this results in very high streamwise vorticity localised at the transition between strips. Analyses not reported here show that the wall-normal location of the spanwise velocity peak scales in inner units when the Reynolds number is increased, while its magnitude becomes $\Rey_\tau$ independent. 

From a qualitative viewpoint, the flow structure predicted by the present linearised model resembles previous experimental observations \citep{wangsawijaya2020} and numerical simulations \citep{chung2018, Neuhauser_Schafer_Gatti_Frohnapfel_2022}. One aspect of discussion concerns the spanwise location of the streamwise-aligned vortices with respect to the alternating pattern of roughness. In the present case, the vortices are symmetrically located above the interface between the strips. This is because the linearity of the governing equations preserves the symmetry that exists across the jump. By contrast, the experiments of \citet{wangsawijaya2020} show that the centres of the vortices are typically found over the low-roughness region. The same was predicted by the simulations of \citet{chung2018}, using an inhomogeneous shear-stress boundary condition to model the roughness. The puzzling aspect is that one would initially attribute the displacement of the vortices to nonlinear convective effects not captured by the linear model, as if the vortices were transported towards the low-roughness region by the relatively intense spanwise velocities near the wall associated to the streamwise vorticity field. However, such a displacement is not observed in the simulations of \citet{Neuhauser_Schafer_Gatti_Frohnapfel_2022} who applied a Navier slip boundary condition for the spanwise velocity to model the roughness, or occurs in the opposite direction in more-recent roughness-resolving simulations of submerged roughness strips \citep{Frohnapfel_von_Deyn_Yang_Neuhauser_Stroh_Gatti_2024}.

From a quantitative viewpoint, a comparison with published results is slightly less straightforward given the particular setup considered in this paper. For this purpose, we use the channel-flow simulations of \citet{chung2018}, at $\Rey_\tau = 590$. The roughness strips are modelled by setting the shear stress to 50\% more and 50\% less than the average shear stress. Assuming that the low and high roughness strips correspond to the smooth and rough wall regions, respectively, \citet{chung2018} estimates that the equivalent sandgrain roughness of the roughness patches is $k_s^+ = 205$. With such settings, they observe maximum wall-normal velocities that peak {between $0.3u_\tau$ and $0.4 u_\tau$} {(see their figure 9a)}. To match the roughness properties of these simulations, one needs to recall that the solution produced by the present linear model is defined per unit variation of the equivalent sandgrain roughness between the strips. From the results of figure \ref{fig:topology}, maximum velocities on the order of $17 \times 10^{-4} u_\tau$ are obtained, for the optimal width $S$. Multiplying this value by $k_s^+ = 205$ we obtain velocities on the order of $0.35 u_\tau$, in very good quantitative agreement with the numerical simulations. Given that the intensity of the cross-stream velocity components characterises somehow the equilibrium between {source and sink} mechanisms of the streamwise vorticity balance \citep{stroh2016, castro2024}, the favourable agreement with simulations suggests that such mechanisms are correctly captured by the linearised RANS model. However, the maximum wall-normal velocity in \citet{chung2018} is obtained for a relatively wide strip, $S = 1.57$, while the present model indicates that the peak occurs at $S\approx 0.7$, and lower velocities are observed for $S=1.57$. It is argued that this is not a Reynolds number effect, but it is due to the vortices in \citet{chung2018} being, as discussed, closer to each other than the strip width $S$ would suggest, resulting in larger induced velocities. Evidence for this is given by the fact that the wall-normal velocities depend on the duty cycle $DC$, as shown later in figure \ref{fig:effect_of_DC}, and peak when the vortices are artificially pushed together by a narrow low-roughness strip, $S_L < S_h$.

\subsection{High- and low-momentum pathways}
A unique characteristic of flows over heterogeneous surfaces is that the longitudinal secondary currents are flanked by high- and low-momentum regions, produced by the vertical ``pumping'' of high and low momentum fluid, respectively, induced by the vortical motions \citep{mejia2013, willigham2014, barros2014}, which determine a significant spanwise alteration of the boundary layer depth. Although a wall-bounded flow above an homogeneous rough surface displays a positive deficit of the streamwise velocity due to the flow deceleration induced by the surface roughness \citep{jimenez2004}, the co-location between the high and low momentum regions and the roughness strips is counter-intuitive as faster flow can be found for certain conditions above the high-roughness strips. In this section, we demonstrate that the present framework clearly captures this phenomenon.

\begin{figure}
    \centering
    % \includegraphics[width=0.9\textwidth]{strips_topology_comparison_ver2.eps}
    %Davide se vuoi rifaccio anche la figura 10 per alcuni casi con DC=0.5 ed alcuni S
    \includegraphics[width=\textwidth]{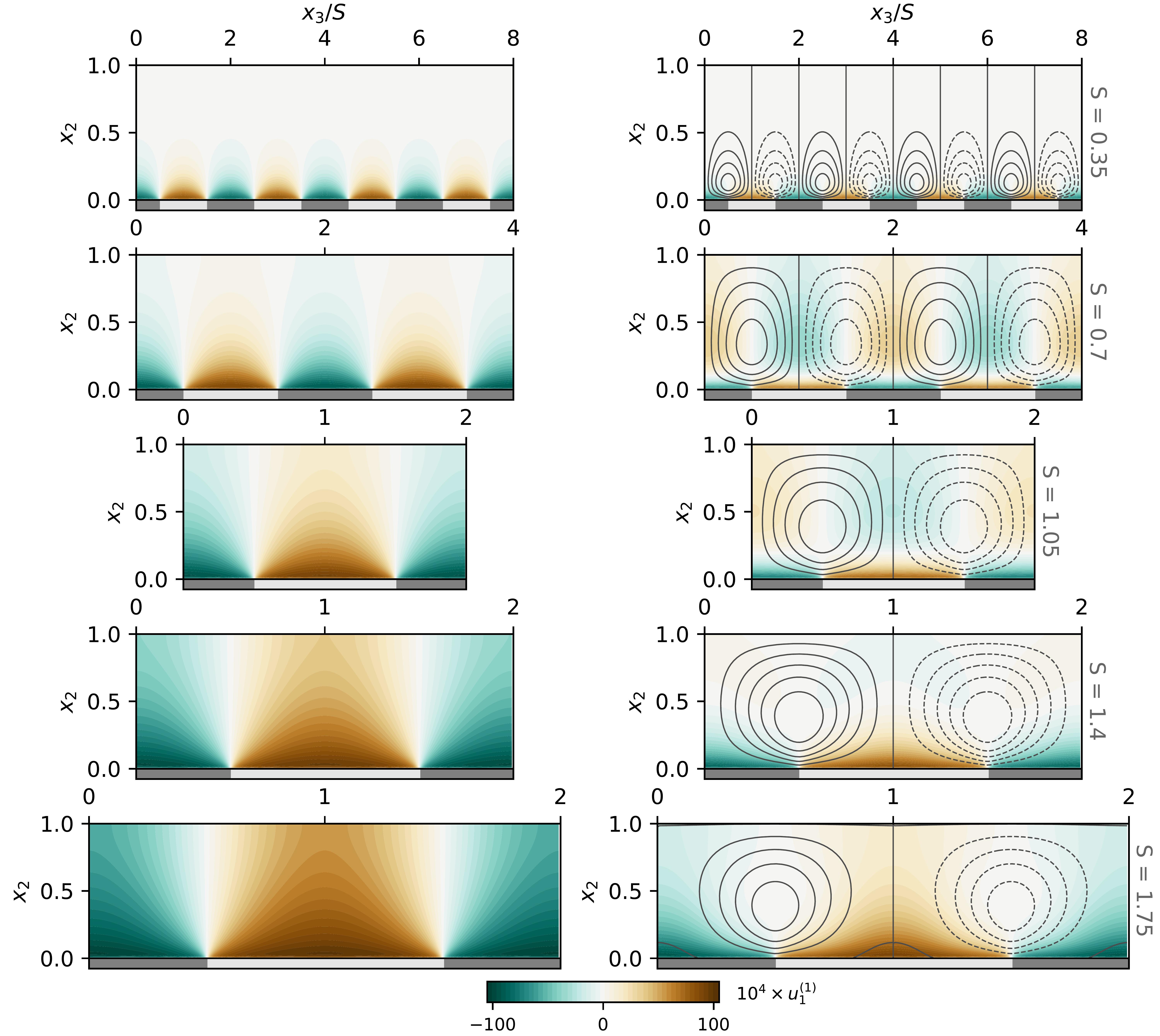}
    \caption{Contours of the streamwise velocity perturbation $u_1^{(1)}$ over roughness strips at $DC=0.5$ for several strip widths $S$, without QCR model (left panels) and with QCR model (right panels). 
    The field in the fundamental domain (see figure \ref{fig:sketch}) is repeated four and two times for the two narrowest strip cases.
    Contour lines of the perturbation of the streamfunction $\psi^{(1)}$ are also reported {to} visualise the secondary flows. Negative $\psi^{(1)}$ contours are indicated by dashed lines. The Reynolds number is $\Rey_{\tau}=1000$, for {$k_s^{(0)} = 180$}. The darker/lighter patches denote the location of the high/low roughness strips.}
    \label{fig:u1_comparison}
\end{figure}

% \begin{figure}
%     \centering
%     % \includegraphics[width=\textwidth]{u1_velocity_strips.eps}
%     \includegraphics[width=1.05\textwidth]{profiles-u1.jpg}
%     \caption{Streamwise velocity component perturbation at the centre of the high-roughness strips for $c_{r1}=0$ (panel a) and $c_{r1}=0.31$ (panel b). The low-roughness width is $W_l=0.67$ while $W_h$ changes from 0.25 to 2 (see the legend). The Reynolds number is $\Rey_{\tau}=1000.$}
%     \label{fig:u1_comparison}
% \end{figure}

% The effect of the QCR nonlinear stress model on the streamwise velocity profile is finally investigated to better understand the generation mechanism of the secondary flows. 
% The deformation of the velocity profiles is hence studied and related to the QCR model. 

In figure \ref{fig:u1_comparison}, contours of the streamwise velocity perturbation $u_1^{(1)}$ are shown for the same configurations of figure \ref{fig:topology}. This visualisation differs to what customarily reported in previous work in that it shows the \emph{velocity deviation} from the streamwise velocity distribution $u_1^{(0)}$ observed over the homogeneous surface. 
% In figure \ref{fig:u1_comparison}, profiles of the streamwise velocity perturbation obtained at the centre of the high-roughness strip $u_1^{(1)}(x_2,0)$ are shown for the same configurations of figure \ref{fig:topology}, shown by the square markers in figure \ref{fig:maps}, i.e.~for increasing $W_h$.
The difference between the maps in the two columns is that in the left panels secondary currents were  artificially ``turned off'' by setting the constant $c_{r1}$ in the nonlinear stress model to zero, so that spanwise and wall-normal gradients of the mean streamwise velocity do not produce any of the Reynolds stresses in the streamwise vorticity equation necessary to sustain longitudinal vortices. In the right panels, solutions for the standard value $c_{r1} = 0.3$ are reported. Without secondary currents, the flow experiences a net deceleration above the high-roughness strips, especially in the near wall region. However, further away from the wall, e.g.~at the centre of the channel, the change in streamwise velocity depends strongly on the strip width, because this parameter controls the depth at which the roughness-induced deceleration ``diffuses'' in the shear flow from the wall due to the turbulent viscosity field. For narrow strips, the velocity deficit at the channel centre is small, and only becomes significant when the the strip width is at least the half the half-height of the channel.  When secondary currents are ``turned back on'', right panels of figure \ref{fig:u1_comparison}, the streamwise velocity over the high-roughness strips is now generally positive because of the down-welling motion in this region, and vice-versa on the low-roughness strips. This only applies for $x_2 \gtrsim 0.1$, because nearer to the wall the local roughness properties control whether the flow is faster/slower. These motions produce dispersive stresses, e.g. $u_1^{(0)}u_2^{(0)}$, that alter the equilibrium in the streamwise direction and result in a non-trivial dependence of the streamwise velocity from the wall. This, fundamentally, implies that the logarithmic velocity distribution is significantly altered by the addition of the dispersive stresses. However, as the strips become wider, secondary currents are not intense enough to produce any significant alteration of the streamwise momentum equilibrium and the deceleration effect produced by the high roughness begins to dominate, starting from the region closest to the wall. Overall, this is the same behaviour observed experimentally \citep{wangsawijaya2020, Frohnapfel_von_Deyn_Yang_Neuhauser_Stroh_Gatti_2024}, where a downward/upward bulging of the contours of the streamwise velocity are observed at the edge of the boundary layer / near the wall.

\begin{figure}
    \centering
    \includegraphics[width=\textwidth]{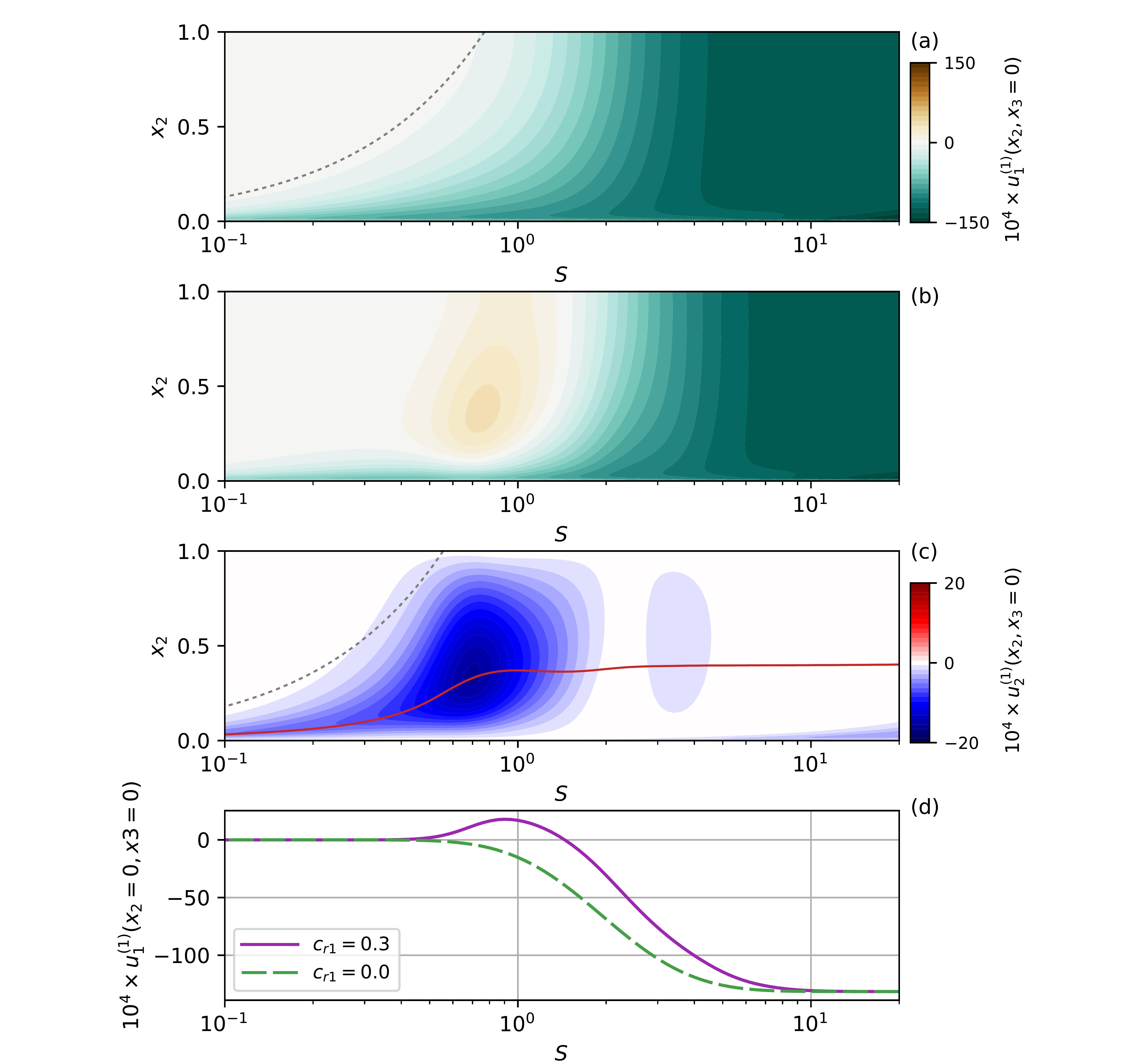}
    \caption{Streamwise velocity extracted on a vertical line at the centre of the high-roughness strip for increasing strip width $S$, for $c_{r1}=0$, (a), and the standard value 0.3, (b). Wall-normal velocity extracted on the same vertical line for $c_{r1}=0.3$, (c). The dashed lines indicates wall-normal distances growing linearly with $S$ for $DC=0.5$. Streamwise velocity perturbation above the centre of the high-roughness strip, at the mid-plane, for the same two cases as a function of the strip width, (d). The Reynolds number is $\Rey_{\tau}=1000$ and {$k_s^{(0)} = 180$}.}
    \label{fig:delta_u}
\end{figure}

The influence of the strip width on the perturbation of the streamwise velocity field is summarised in figure \ref{fig:delta_u}. The streamwise velocity profile at the centre of the high-roughness strip is extracted from several calculations with $S$ in the range $[0.1, 20]$, with and without the nonlinear Reynolds stress model. These profiles are concatenated together to form the maps in panels (a) and (b), respectively. For the case with $c_{r1}=0.3$, we also report a similar plot for the wall-normal component, in panel (c). Given that a logarithmic velocity shift is not a meaningful quantity to compute, we report in panel (d) the velocity deviation in the centre of the channel, above the centre of the high-roughness strips. In panel (c), the solid red line denotes the wall-normal location of the perturbation streamfunction peak.
The three regimes discussed in \citet{chung2018} can be clearly identified. For wide strips, the velocity deficit tends to a value controlled by the roughness function at $k_s^{(0)}$ for both cases, as the turbulent structure over each strip tends to its equivalent over an homogeneous surface given that the influence of neighbouring patches and of the secondary currents localised at the transition between strips vanishes. In this regime, the model predicts that the centre of the rolls is located at a distance from the wall of about $0.42$, i.e. the rolls are space-filling in the vertical direction. The streamwise velocity perturbation is also roughly constant as a function of the distance from the wall for $S \gtrsim 8$, producing the expected shift of the logarithmic velocity profile over each strip, regardless of whether the nonlinear Reynolds stress is active or not.

For narrow strips, the height of the channel is much larger than the ``depth'' at which the effect of the wall inhomogeneity is perceived, analogously to the blending height concept for spatially varying roughness discussed in \citep{bou2007parameterization}. For the case $c_{r1}=0$, without rolls, this depth (estimated from the contours of the streamwise velocity perturbation) varies linearly with $S$, as one would expect to see in a diffusion-driven problem, given that the turbulent viscosity does not vary significantly except for near the wall. For the case $c_{r1}=0.3$, with rolls, the depth measured in terms of the wall-normal velocity also appears to increase linearly with $S$, at least for the smallest $S$ analysed here. However, the linear scaling (the dashed lines) appears to {lose} accuracy relatively rapidly, at $S\approx 0.25$, as soon as the rolls occupy about half of the half-channel height.
% it might be possible to develop a scaling law to express the wall-normal location of the centre of the rolls as a function of $S$, when $S \ll 1$ and, of course, when $S^+$ is much large than the roughness height. However, at the relatively low Reynolds numbers considered here, $\Rey_\tau = 1000$, such a scaling law does not clearly appear from the data. In fact, depth at which aa significant wall-normal velocity panel (c) seem to grow as $\sqrt{S}$ for $S\ll 1$, but the fit looses accuracy as $S \rightarrow 1$. 
In between these two regimes, in the ``transitional regime'' of \citet{yang2017}, the streamwise velocity perturbation display a complex behaviour, highlighting the significant lack of flow homogeneity. At $S=0.7$ the rolls are most intense and induce the maximum perturbation at a wall-normal location close to the centres of the rolls. However, the velocity perturbation reaches its peak at the mid plane only at $S=0.95$ and then changes sign from $S \gtrsim 1.75$. Interestingly, the model never predicts flow reversal when the strip width is increased, in agreement with observations in the literature {\citep{Neuhauser_Schafer_Gatti_Frohnapfel_2022}}

Overall, the present framework appears to capture correctly the three flow regimes documented in the literature. The implication is that linear mechanisms, whereby secondary flows may be interpreted as the output response of the mean shear flow to a steady forcing localised at the wall, may be sufficient to predict the size and strength of the rolls. Based on the streamwise velocity evaluated at the mid-plane, the boundaries may be located at $S \approx 0.4$ and $S \approx 8$, but differences may arise with alternative criteria that consider the bulk of the flow.   

\subsection{Eddy viscosity perturbation}
{In many studies that have examined the properties of the linearised Navier-Stokes equations, velocity perturbations resulting from optimal or stochastic forcing are computed by assuming that the turbulent viscosity distribution is not affected by the flow perturbation and follows an analytical or empirical distributions \citep{reynolds1972, delalamo2006, hwangcossu2010, morra_2019, Pickering_Rigas_Schmidt_Sipp_Colonius_2021}}. In the present case, and unlike previous work, the governing equations must include a transport equation for the turbulent viscosity. {It is speculated that the peak location predicted by the present linear model (see figure \ref{fig:maps}) is determined in large part by the {selectivity} of the linearised Navier-Stokes operator, rather than by the specifics of the turbulence model adopted}. However, the inclusion of such a model and the resulting perturbation of the turbulent viscosity $\nu_t^{(1)}$ is key to capture the influence of the heterogeneous surface roughness and the generation of secondary flows, as described in section \ref{sec:turbulence modelling}. In practice, spanwise gradients of the eddy viscosity produce, through the QCR model, Reynolds stresses that act as source terms for the streamwise vorticity equation, resulting in secondary motions. 

\begin{figure}
    \centering
    \includegraphics[width=\textwidth]{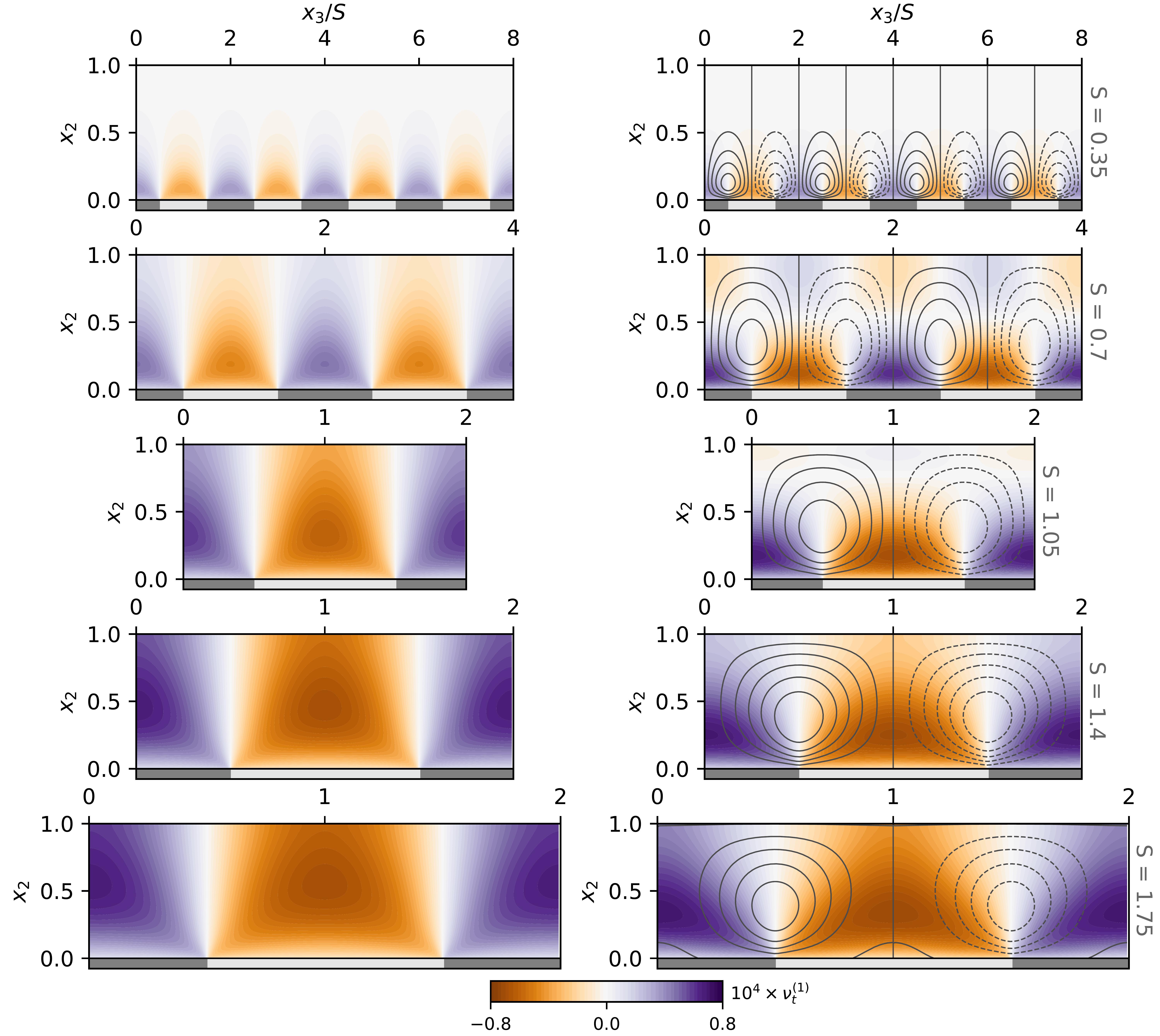}
    \caption{Contours of the eddy viscosity perturbation $\nu_t^{(1)}$ over roughness strips at $DC=0.5$ for several strip widths $S$, without QCR model (left panels) and with QCR model (right panels). 
    The field in the fundamental domain (see figure \ref{fig:sketch}) is repeated four and two times for the two narrowest strip cases.
    Contour lines of the perturbation of the streamfunction $\psi^{(1)}$ are also reported {to} visualise the secondary flows. Negative $\psi^{(1)}$ contours are indicated by dashed lines. The Reynolds number is $\Rey_{\tau}=1000$, for {$k_s^{(0)} = 180$}. The darker/lighter patches denote the location of the high/low roughness strips.}
    \label{fig:nut_comparison}
\end{figure}

Figure \ref{fig:nut_comparison} shows contours of the perturbation turbulent viscosity for several strip widths $S$. In the left panels, the QCR constant $c_{r1}$ has been set to zero, to characterise the eddy viscosity distribution in the absence of cross-stream motions and highlight the interaction with the streamwise velocity fields of figure \ref{fig:u1_comparison}. In the right panels, the QCR constant is $c_{r1} = 0.3$. In general, positive eddy viscosity perturbations are observed above the high roughness strips, and vice versa, reflecting the boundary condition \eqref{eq:BCs_first} and the altered distance $d$ from the wall. For narrow strips, the eddy viscosity perturbation is confined near the wall, given that rapid spatial variations of the eddy viscosity tend to be damped by the diffusion term in the linearised SA equation. As the strip gets wider, more intense eddy viscosity distributions are observed, reflecting the increased acceleration/deceleration of the flow over the low and high roughness strips, respectively.
Small, but likely significant, changes are observed when secondary flows are ``turned on'', {see the right panels of figure \ref{fig:nut_comparison}}. It can be observed that the cross-stream motions produce a further distortion of the eddy viscosity distributions. {This is the result of two mechanisms that result from the analysis of the SA transport model: a) the advection of turbulent viscosity of the background flow operated by the vertical and lateral velocities and b) the altered production of turbulent viscosity due to the alteration of wall-normal and spanwise gradients of the streamwise velocity.}

\subsection{Tertiary structures and role of the duty cycle}
\begin{figure}
    \centering
    \includegraphics[width=\textwidth]{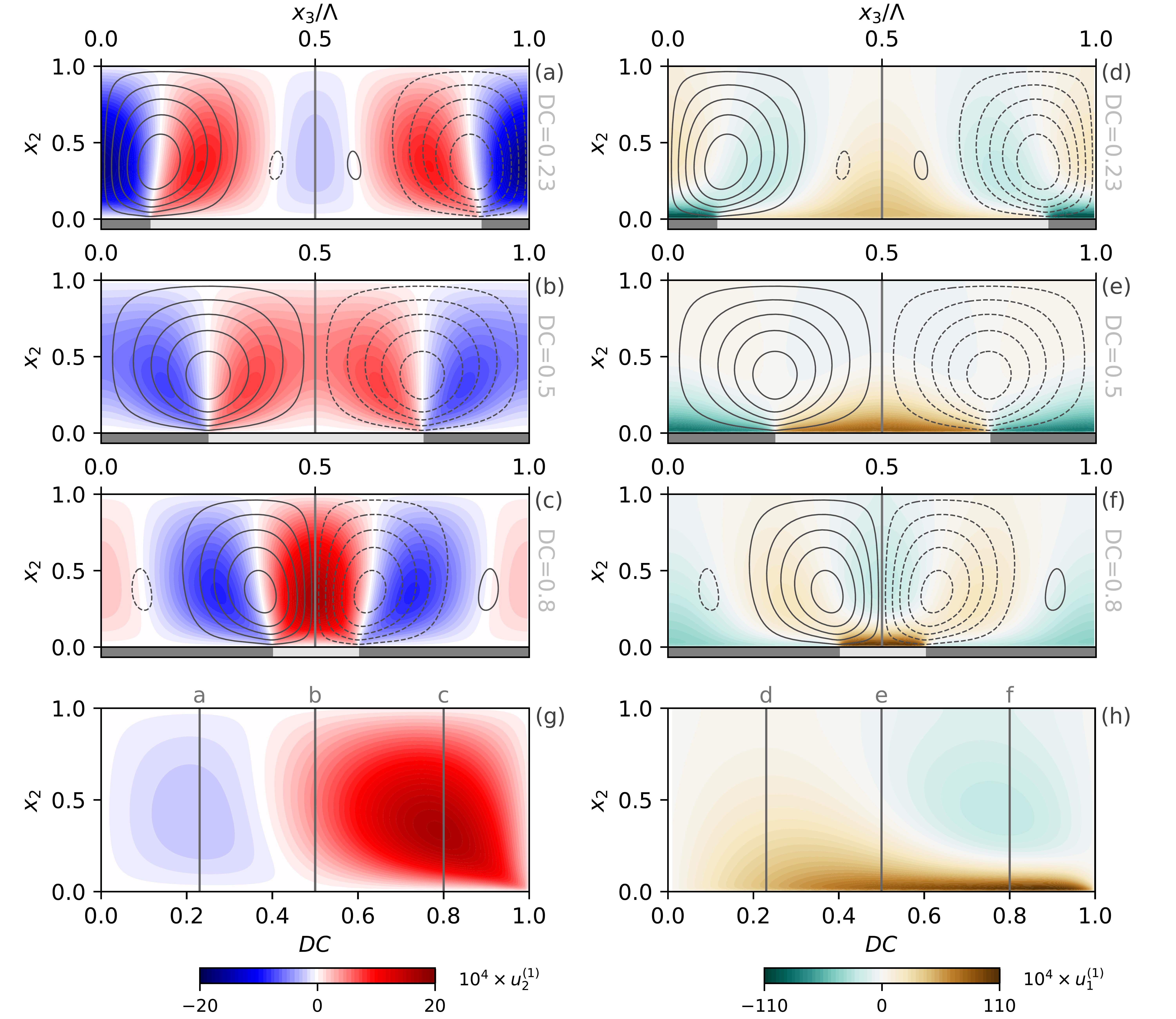}
    \caption{Contours of the wall-normal (a-c) and streamwise (d-f) velocity perturbation for increasing duty cycles, indicated in the figures, and $\Lambda=2.8$. Panels (g) and (h) show the wall-normal and streamwise velocity, respectively, extracted on the vertical line (shown in panels (a-f)) at the centre of the low-roughness strip $x_3/\Lambda = 0.5$ for duty cycles ranging from 0 (narrow high-roughness strip) to 1 (wide high-roughness strips). The lines labeled (a-f) in panels (g, h) refer to the corresponding panels. The Reynolds number is $\Rey_{\tau}=1000$ and {$k_s^{(0)} = 180$}.}
    \label{fig:effect_of_DC}
\end{figure}
% For wide strips, the model does not predict a coherent flow reversal for $DC=0.5$, but significant alterations of the flow structure above the high-roughness region can be found 
% for $W_h \approx 2$, as tertiary flows form reversing the flow direction over the high-roughness strip. Spanwise velocity profiles, taken at the interface between the high and low roughness strips are displayed in panel (b) of figure \ref{fig:profile}. 
% and eventually weak tertiary flows occur at the centre of the high-roughness regions. These turbulent structures have been observed for ridge-type roughness and they emerge in the gap between ridges when the distance between individual ridges is large enough, or at the centre of wide ridges \citep{vanderwel2015, medjnoun_vanderwel_ganapathisubramani_2020}. 
The present model shows that the duty-cycle appears to play an important role in controlling the formation of tertiary structures, which have so far not been observed for strip-type roughness \citep{Neuhauser_Schafer_Gatti_Frohnapfel_2022}. This is shown in figure \ref{fig:effect_of_DC}, where contours of the wall-normal and streamwise perturbation velocities are shown for increasing duty cycles, at $\Lambda=2.8$. These cases correspond to the triangular marker in figure \ref{fig:maps}. For each of these solutions, as well as for other solutions in the interval $DC = [0, 1]$, the wall-normal and streamwise velocity profiles are extracted at the centre of the low-roughness strip. All these profiles are then combined together to form the colour map reported in the bottom panels of figure \ref{fig:effect_of_DC}. This visualisation suggests that flow reversal over the low-roughness strip may begin to appear in practice as the duty cycle is decreased to about $0.4$ and would be the most intense for $DC \approx 0.23$ (i.e. for $S_h \approx 0.3 S_l$). Incidentally, the occurrence of flow reversal and tertiary structures explains the two ``ears'' of the contours of the kinetic energy density map of figure \ref{fig:maps}(a), where the cross-stream motions are relatively intense, compared to other duty cycles. This analysis also suggests that the vertical velocities over the low-roughness strip are most intense for a duty cycle equal to about 0.8 and not for the symmetric case $DC=0.5$ which the kinetic energy density maps would suggest. Arguably, this can be attributed to the constructive interference of the wall-normal velocities induced by two neighbouring vortices, pushed closer to each other by the decreasing width of the low-roughness strip. The streamwise velocity field is also particularly affected by the duty cycle, as faster or slower flow over the low roughness strip can be found depending on the $DC$. The transport of fast/slow fluid operated by the cross-stream velocities is particularly visible. For instance, for $DC=0.23$ the left vortex, rotating counter-clockwise transports low-momentum fluid from the near wall region at $x_3 = 0$ to its right flank, producing a negative velocity streak at $x_3 \approx 0.2$, and similarly for the other longitudinal vortex.

\section{Generalising the framework to complex surface heterogeneities}\label{sec:combination}
It has been demonstrated that when conducting experiments \citep{wangsawijaya2020} or roughness-resolving simulations \citep{stroh2019} over realistic heterogeneous rough surfaces it is {pivotal} to ensure that the shear-increasing effects of roughness are decoupled from the inevitable variation of the mean surface height. Both roughness and elevation heterogeneity produce secondary currents and therefore the combination of such effects can significantly influence the strength and potentially the direction of the resulting secondary motions \citep{schafer2022, Frohnapfel_von_Deyn_Yang_Neuhauser_Stroh_Gatti_2024}. In this section, we analyse this aspect through the lens of the linearised model, to initiate the formulation of a unifying framework for flows over complex heterogeneous surfaces. 

\subsection{Secondary-flow-inducing source mechanisms}
In the present linearised framework, in the limit case where the spanwise variation of the roughness or the elevation is small, ridge-type and strip-type roughness are modelled with the same approach. {In both cases, the flow-surface interaction develops through three separate source mechanisms corresponding to three different inhomogeneous terms acting as forcing in the linearised equations. To illustrate these mechanisms, it is instructive to examine ridge-type roughness considered in our previous work \citep{zampino2022} where the mechanisms are all active.}
% suitable wall boundary conditions
% to asses the impact of the lateral variation of the mean roughness height and to 
% There are important similarities between models for surfaces with alternating roughness strips and for surfaces with longitudinal ridges \citep{zampino2022}  
% The only difference in how these two types of heterogeneity are modelled lies in the boundary conditions 
% on the streamwise velocity and the eddy viscosity, as the governing equations are exactly the same. 
In such case, the lateral variation of the elevation was defined by a unitary peak-to-peak, zero-mean function $f(x_3)$ so that, e.g., the bottom wall of the channel is located at $x_2 =  \epsilon f(x_3)$ and the small parameter $\epsilon$ controls the actual amplitude of the topography. The first source mechanism, denoted as $A$ in what follows, is mediated by the linearised boundary condition on the streamwise velocity, e.g. on the lower wall,
% \begin{subeqnarray}
\begin{equation}
    u_{1}^{(1)}(x_2 = 0) = -  f(x_3) \left.\frac{\partial u^{(0)}}{\partial x_2}\right|_{x_2= 0 } = -  f(x_3) \Rey_{\tau},
\end{equation}
{derived in \citet{zampino2022}.}
Physically, this condition produces a velocity slip that captures the acceleration and deceleration perceived by the bulk flow above the troughs and the crests of the non-planar topography, respectively. What leads to the formation of secondary flows is the resulting spanwise gradient of the streamwise velocity in the near wall region. This gradient induce, via the QCR model, spanwise gradients of anisotropic Reynolds stresses {(see equation \eqref{tau23qcr})} that then create secondary flows. This mechanism does not seem to have been discussed previously in the literature of secondary flows. More generally, we are not aware of studies that consider the somewhat artificial surface arrangement consisting of longitudinal, flush-mounted belts moving upstream/downstream alternated with regions of solid wall. This setup would capture source mechanism $A$ directly, and we predict that it could generate relatively intense secondary currents with a down-welling over the upstream-moving belts. When modelling strip-type roughness, this source mechanisms is not active since no-slip boundary conditions for the streamwise velocity are used. This is equivalent to stating that the mean height of the two roughness strips is the same and the boundary of the numerical domain is at some suitable location where the strip-averaged streamwise velocity goes to zero.

The second source mechanism, denoted as $B$, {is active for both types of heterogeneity. It} is mediated by the destruction term of the SA transport equation, where the inverse of the squared distance $d$ between a point in the numerical domain and the nearest ``wall'' models the blocking effect of the wall \citep{spalart1994, aupoix2003}. With such term, the SA model {predicts} an accurate log-layer, and thus lateral perturbations of the distance $d$ {induced} by the elevation (via the function $f(x_3)$) or by the displacement of the virtual origin (via the term $d_0^{(1)}(x_3)$), produce a {perturbation} of the the log-layer and spanwise gradients of the turbulent viscosity field. Crucially, the sign of this source mechanism is opposite for strip-type and ridge-type roughness: while locally increasing elevations correspond to locally reducing distances $d$, increasing roughness produces  increasing distances, as the virtual origin is further displaced downwards beneath the boundary of the numerical domain. In this regard, the framework suggests that the mean roughness height is not the important factor. Rather, the displacement of the virtual origin, a dynamic parameter that ultimately depends on the drag of the surface, is what controls the intensity of the forcing and of the resulting secondary flows. As a side note, while source mechanism $A$ acts as a boundary condition, source mechanism $B$ acts at all wall distances, as it captures the perturbed development of the wall-bounded flow from a different origin.

The third source mechanism, denoted as $B^\prime$, is {again active for both types of heterogeneities. It is} mediated by the inhomogeneous boundary condition on the modified turbulent viscosity. For ridge-type roughness, this is
\begin{equation}\label{eq:bc_nu_fx3}
    \tilde{\nu}^{(1)} (x_2= 0) = - f(x_3) \left. \frac{\partial \tilde{\nu}^{(0)}}{\partial x_2}\right|_{x_2=0} =-f(x_3) \kappa.
\end{equation}
For strip-type roughness, the lateral variation of the virtual origin $d_0^{(1)}(x_3)$ plays the same role of the function $f(x_3)$ describing the ridge topography, but with an opposite effect as dictated by the boundary condition \eqref{eq:BCs_first}. This condition must be applied consistently with source mechanism $B$, so that the shifted eddy viscosity profile produced by such mechanism is consistent with the boundary condition \eqref{eq:bc_nu_fx3}. Studies on modelling roughness in RANS simulations \citep{aupoix2003} have shown (and we confirm it later) that this  third mechanism is quite weak, because capturing the overall development of the turbulent structure from a different virtual origin is more important than applying non-zero boundary conditions for the turbulent quantities.

\subsection{Combining elevation and roughness variations}
\begin{figure}
    \centering
    \includegraphics[width=\textwidth]{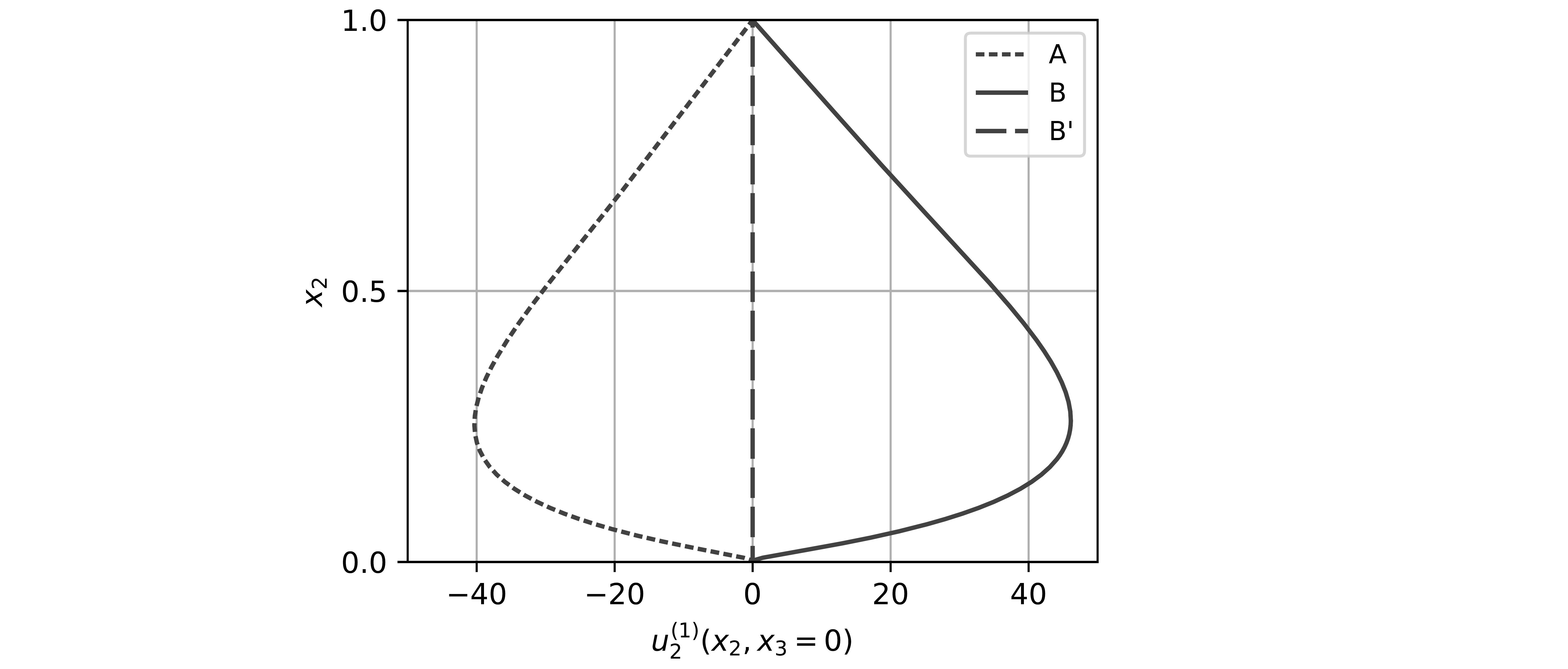}
    \caption{Wall-normal velocity profiles over the crest of a sinusoidal topography, for $\Lambda=1$ and $\Rey_\tau = 1000$, obtained by activating each of the three secondary-flow-inducing source mechanisms in turn.}
    \label{fig:combination-of-inhomogeneities}
\end{figure}

To examine the relative strength and the combination of these three source mechanisms, we first perform linearised calculations for $\Rey_\tau = 1000$ for a smooth sinusoidal wall where \mbox{$f(x_3) = \cos(2\pi / \Lambda x_3)$}, with period $\Lambda = 1$, by activating only one source mechanism each time. Results are reported in figure \ref{fig:combination-of-inhomogeneities}. The wall-normal velocity profiles, taken at $x_3=0$ over the crest of the topography, show that source mechanism $A$ produces a down-welling flow over the region where the slip velocity is negative. Conversely, the decrease of the distance from the wall over the crest, source mechanism $B$, produces an up-welling of slightly greater magnitude, while source mechanism $B^\prime$ is much weaker than the first two, as discussed. 
Because the proposed model is linear, the superposition principle applies and the effect of varying simultaneously the roughness properties and the elevation can be obtained easily by combining appropriately solutions obtained in the two cases. For ridge-type roughness, source mechanisms $A, B$ and $B^\prime$ 
are all active, while for strip-type roughness only source mechanisms $B$ and $B^\prime$ should be retained, after inverting the sign of the induced flow given the different orientations of $f(x_3)$ and $d_0^{(1)}$ (see figure \ref{fig:combination}).

% can be superimposed to obtain the wall-normal velocity profile in panel (b). Both source mechanisms $A$ and $B$ are relatively intense, but source mechanism $B$ is slightly dominant and an up-welling is predicted over the crest of the sinusoidal topography. However, the net wall-normal velocity, i.e. the intensity of the secondary currents, is much weaker than one would expect from the relative strengths of the two source mechanisms, because of the ``damping'' effect produced by source mechanism $A$ and the associated spanwise variation of the slip velocity.  By contrast, insight into the flow structure that would be generated by the above mechanisms over strip-type roughness can be obtained by only retaining source mechanisms $B$ and $B^\prime$, and inverting the sign of the induced flow given the different orientations of $f(x_3)$ and $d_0^{(1)}$ (see figure \ref{fig:combination}). The wall-normal velocity in now negative where the virtual origin has been displaced downwards, in the region of high roughness. Importantly, it is much more intense than the velocity produced by combining source mechanisms $A, B$ and $B^\prime$.

\begin{figure}
    \centering
    \includegraphics[width=0.999\textwidth]{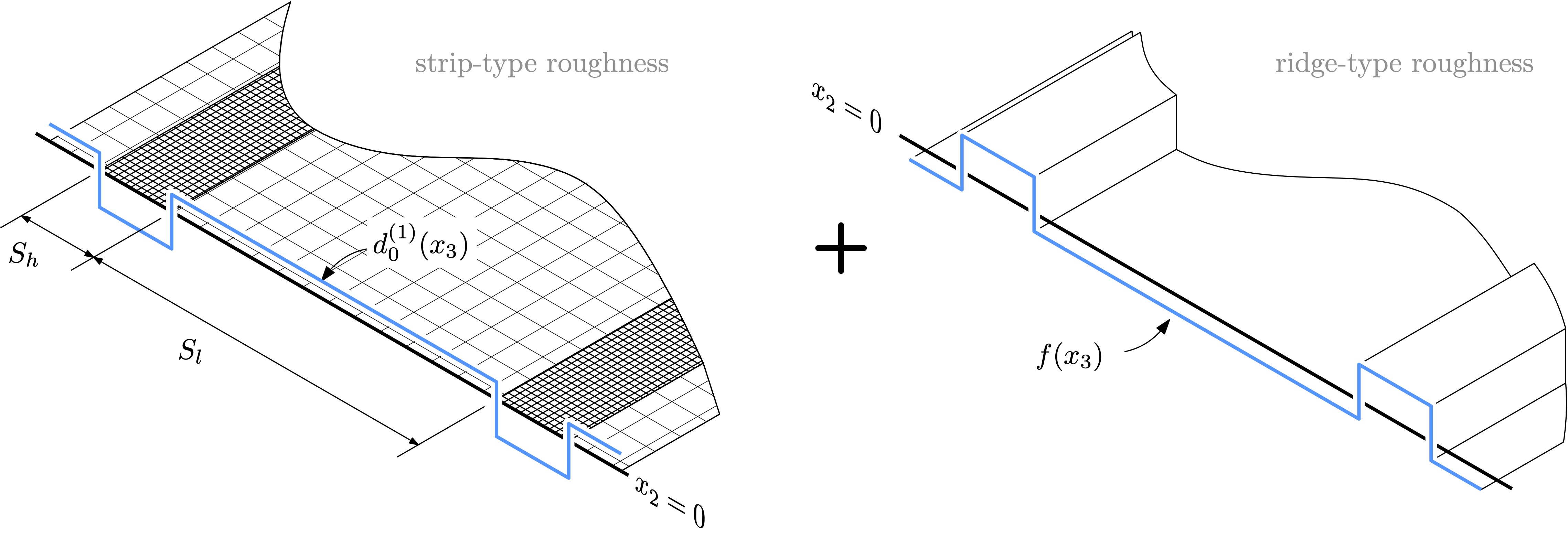}
    \caption{Superposition of high and low roughness strips with smooth rectangular ridges modelling the effect of lateral variation of elevation over the bottom wall of the channel. The lateral variation of the virtual origin of the rough surface is captured by $d_0^{(1)}(x_3)$ (positive if the virtual origin is below the mean height at $x_2=0$), while the lateral variation of the elevation is captured by the function $f(x_3)$ (positive if it is above $x_2=0$).}
    \label{fig:sketch-combination}
\end{figure}

\begin{figure}
    \centering
    \includegraphics[width=\textwidth]{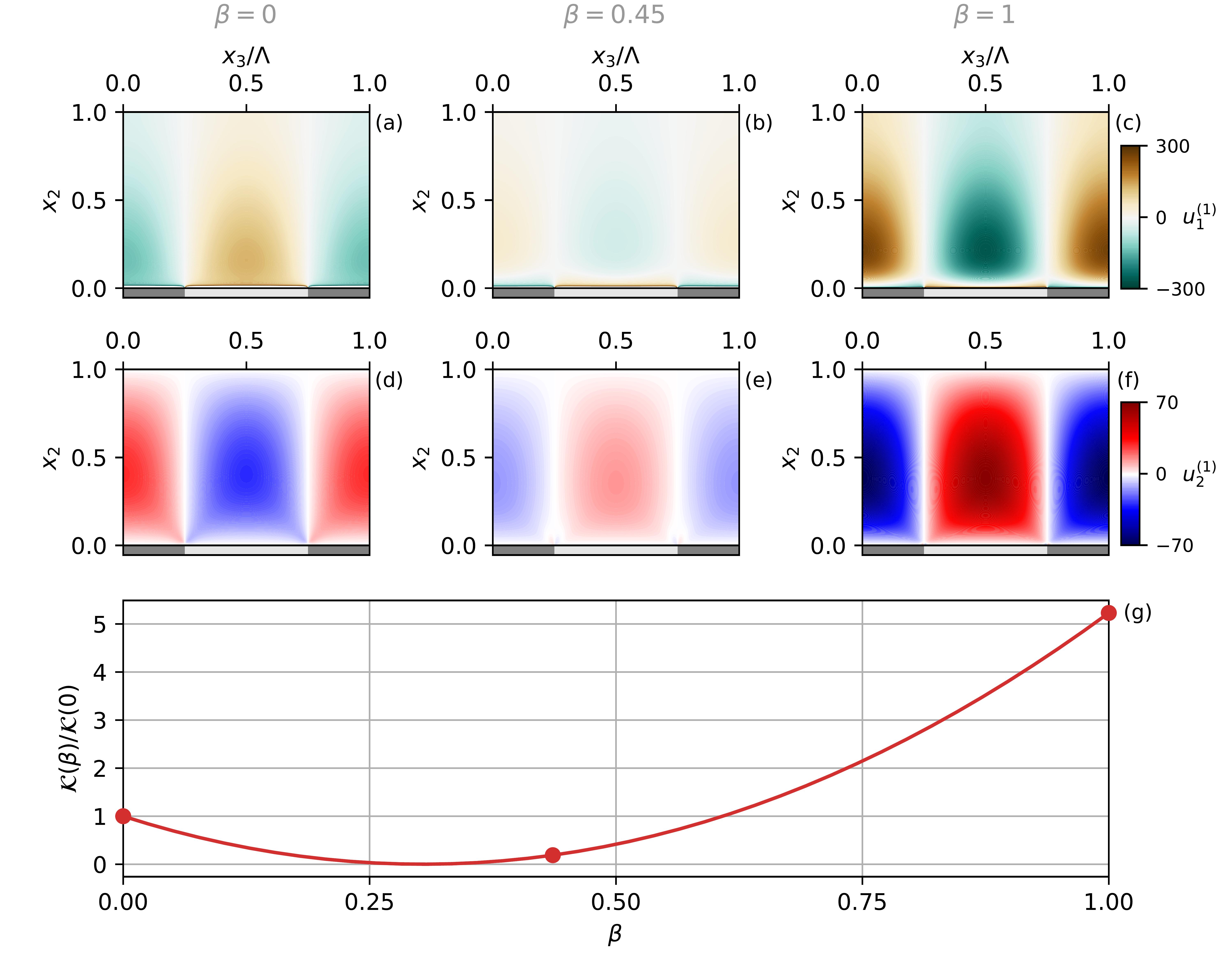}
    \caption{Contours of the perturbation streamwise (a-c) and wall-normal (d-f) velocity components for combinations of ridge-type and strip-type roughness for $S_h = S_l = 0.7$. The parameter $\beta$ is 0.0, 0.45 and 1, as we move from left to right with configurations where the effect of the elevation heterogeneity dominates to cases where the roughness heterogeneity prevails. 
    Normalised kinetic energy density of the combination of both ridge-type and strip-type roughness as a function of the parameter $\beta$,  {(f)}. The Reynolds number is fixed at 1000 and {$k_s^{(0)}$ = 180}.}
    \label{fig:combination}
\end{figure}

To better demonstrate the relative importance of these mechanisms, we then consider a configuration where strips and rectangular ridges are arranged in phase, and the high-roughness regions are placed over the ridges (see figure \ref{fig:sketch-combination}). The width $S_l$ coincides with the gap between the ridges while $S_h$ coincides the ridge width (denoted as $G$ and $W$ in \citet{zampino2022}). To characterise the relative strength of the two effects, we introduce the parameter $\beta$ as the ratio between the displacement of the virtual origin produced by the strips and the topography, so that 
\begin{equation}
     d_0^{(1)}(x_3) = \beta f(x_3),
\end{equation}
with the caveat that positive displacements are in different directions depending on the type of heterogeneity. Case $\beta=0$ corresponds to smooth ridges, leading to secondary flows produced by lateral variations of the elevation. As discussed in \citet{zampino2022}, the linearised RANS model predicts an upwelling over the high-elevation regions. On the other hand, case $\beta=1$ corresponds to the combination of the two roughness heterogeneities where the downward displacement of the virtual origin produced by the shear-increasing roughness is, in theory, fully compensated by a increased elevation.

Figure \ref{fig:combination} shows results of this analysis, where we conduct calculations over surfaces with combined roughness and elevation as a function of $\beta$. The kinetic energy density obtained at each composite surface is normalised with the reference value at $\beta=0$ and is shown in panel (g). Results are shown for $\Rey_\tau = 1000$ and {$k_s^{(0)} = 180$}, for strip (and ridge) widths configurations characterised by constant $S_h = S_l = 0.7$.
% and for the configuration $W_h=W_l=0.67$. In panel (a), the maps for constant $W_l=0.67$ and varying $W_h$ is displayed for $\beta$ between 0 (ridges with smooth surface) to 1 (ridges with rough surface) in order to investigate the effect of $\beta$ on the strength of these vortices. 
The kinetic energy shows a minimum for $\beta\approx0.3$, where the cross-stream velocity components vanish and no secondary flows are predicted. The location of the minimum does not seem to depend greatly on the strip configuration {(not shown for brevity)}. For $\beta > 0.3$ the effect of the lateral variation of the roughness becomes dominant and the associated kinetic energy density can be several times higher than the reference value. 
% \begin{figure}
    % \centering
     % \includegraphics[width=0.55\textwidth]{Topology_article_K_wk.eps}
    %\includegraphics[width=\textwidth]{kinetic_combination.eps}
    % \caption{
    % \label{fig:combination}
% \end{figure}
%The comprehensive map is finally displayed in figure \ref{fig:map_beta} for $W_h=W_l$ and varying $\beta$. It is worth noting two peaks: (i) at $\beta=0$ and $W_h=W_l=0.67$ corresponding to the peak amplification above smooth rectangular ridges as reported in \citet{zampino2022} and (ii) at $\beta=1$ and $W_h=W_l=0.67$ for the rough ridges.
% \begin{figure}
%    \centering
%    \includegraphics[width=0.6\textwidth]{Topology_article_combination_map_wk_contour.eps}
%    \caption{Colormap of $\mathcal{K}_\beta/\mathcal{K}_0$ at varying $\beta$ and $W_h=W_l$. The friction Reynolds number is $\Rey_\tau=1000$. Isolevels for $\mathcal{K}_\beta/\mathcal{K}_0$ from 0.01 and 5 are also plotted for completeness.}
%    \label{fig:map_beta}
% \end{figure}
% This is not the case of the present analysis because the minimum occurs for $\beta=0.3$. The reason is These conditions imposed to model the ridge-type roughness largely modify the flow organisation and introduces an additional forcing to the generation of the secondary flows.  In fact, when removing the boundary conditions for $u_1^{(1)}$ $\mathcal{K}_{\beta=1}$ tends to zero.  
%Che tipo di equazioni?
%(see figure da aggiungere) 
The resulting flow structure as $\beta$ is increased, i.e. as the effect of roughness is increased, is shown in panels {(a-f)}. Maps of the perturbation streamwise and wall-normal velocity components are shown, for $\beta=0$, in the region where the ridge-type roughness is dominant, $\beta=0.45$ close to the minimum of $\mathcal{K}(\beta)/\mathcal{K}(0)$, and 1, where the strip-type roughness is dominant. For $\beta=0$ (left column), the flow topology shows an upwelling over the ridges, as the upward displacement of the wall produced {by} the ridge-type roughness in this region is dominant. For $\beta=0.45$ (central column), secondary currents have changed in direction but their strength is relatively weak. For $\beta=1$ (right column), the flow topology shows a downwelling over the high roughness strip, as the effect of the downward displacement of the virtual origin produced by the roughness prevails in this region over the influence of the increased elevation.
% similar to what is expected for the strip-type roughness but the peak velocity is stronger as expected from the figure \ref{fig:combination}.

The key result of this analysis is that lateral variations of the roughness properties and of the elevation do not have the same impact on the strength of the resulting secondary motions, because of the ``damping'' effect of the spanwise variation of the slip velocity produced by source mechanism $A$, active for heterogeneous elevation but not for heterogeneous roughness. 
% This would be the case if the minimum of curves in figure \ref{fig:combination} occurred at $\beta=1$, since the displacement of the virtual origin induced by the strips in this case is equal in value but opposite in sign to that induced by the ridges. 
The location of the minimum of $\mathcal{K}(\beta)$ suggests that the secondary-flow-inducing effect of roughness is about three times stronger than that of ridges, for the same displacement of the virtual origin in absolute terms. However, this prediction is clearly no better than the prediction of the strength of secondary flows for the two heterogeneities. The comparison with the heterogeneous roughness simulations of \citet{chung2018} reported earlier suggests that the linear model captures quite well the strength of secondary motions over such surfaces. However, for ridges, recent work \citep{castro2024} has suggested that the linear model over-predicts the intensity of secondary motion for tall ridges, owing to the importance of nonlinear effects near the corners of the ridges, while predictions can be more accurate for short ridges that do not protrude excessively in the wall-bounded flow. Overall, this indicates that the response of the wall-bounded flow to a perturbation of the surface elevation is far from being linear. To our knowledge, evidence for this claim was perhaps first given in \citet{Wang2006} (see their figure 20), who showed how the vertical velocity produced by secondary motions saturates rather quickly as the height of the topography is increased. This last piece of evidence suggests that the secondary-flow-inducing effect of the roughness might be even stronger than what the linear model suggest here, although a precise quantification might require dedicated experimental work. Overall, this could also justify the recent results of \citet{Frohnapfel_von_Deyn_Yang_Neuhauser_Stroh_Gatti_2024} who considered the same surface arrangement considered in this section, with roughness strips located in phase with the ridges. These authors increased the ridge height but did not observe a reversal of the flow direction above the ridges, which was dominated by the down-welling caused by the roughness. Using their data, we calculate the height of the ridges to be 5.16\% the height of the channel and the downward displacement of the virtual origin to be 1.08\%, from their homogeneous roughness data, for a ratio $\beta \simeq 0.21$. This ratio is clearly to the left of the minimum in figure \ref{fig:combination}, reinforcing the idea that the response of the wall bounded flow to finite lateral variations of the elevation are less intense than predicted by a linear model for infinitesimal perturbations.

\section{Discussion and concluding remarks} \label{sec:conclusions}
In this paper, we proposed a linearised-RANS-based framework to predict the structure of Prandlt's secondary flows of the second kind developing over laterally heterogeneous rough surfaces. The work extends our previous efforts on modelling smooth non-planar surfaces \citep{zampino2022}, e.g. surfaces with longitudinal ridges. The model couples the linearised RANS equations with the Spalart-Allmaras (SA) transport equation to capture the altered turbulent structure. Rough surfaces with alternating streamwise-aligned strips of high and low roughness are modelled using established RANS modelling strategies available in the literature for homogeneous rough surfaces \citep{Aupoix2007}. {Briefly, these strategies adopt a virtual origin framework, whereby the shift of the logarithmic profile is obtained by displacing beneath the boundary of the numerical domain the origin from which turbulent quantities develop \citep{rotta1962}}. This results in altered boundary conditions as well as a domain forcing term when the distance from the wall appears in the turbulence model's transport equations. The framework also employs a nonlinear Reynolds stress model, i.e.~the Quadratic Constitutive Relation (QCR, see \citet{spalart2000}), so that secondary currents induced by the inhomogeneity of anisotropic turbulent stresses can be predicted \citep{speziale1982}.

There are several aspects that the model predicts remarkably well in agreement with previous observations. One first aspect is the presence of three separate flow regimes as the strip width is increased \citep{chung2018}. For narrow strips, the linear model {supports} a flow structure consisting of rolls localised in the vicinity of the wall and having wall-normal size scaling linearly with the strip width. For wider strips, the flow structure tends to an ``isolated-vortex'' regime, where streamwise vorticity is concentrated in a roughly square region localised around the transition between strips, while the bulk flow in the centre of the roughness strips tends to its homogeneous rough-wall flow dictated by the local roughness properties. In the intermediate regime, secondary currents are most intense when the high and low roughness strips have the same widths, and are equal to about $0.7$ of the half-channel height. The model also provides adequate quantitative predictions of the intensity of the cross-stream velocity component, compared to, e.g., the numerical simulations of \citep{chung2018}. A second aspect concerns tertiary structures and flow reversal, not observed in previous studies on roughness strips that have most often examined high and low roughness strips of equal width. The linear model predicts that these phenomena only appear when the strips have different width. For instance, for $\Lambda = 2.8$, flow reversal is strongest on the low-roughness strip when this strip is about 4 times wider than the high-roughness strip. It would be interesting to confirm this prediction through experiments or simulations. A third aspect concerns the occurrence of low- and high- momentum pathways flanking the longitudinal rolls, where high-speed flow may be found on the high-roughness strip (and vice-versa) in regions dominated by the vertical velocities induced by the rolls. Away from the rolls, or for wide strips, the expected relationship between surface roughness and streamwise velocity defect is recovered. 

% This is the same width for which the linear model predicts secondary currents with maximum strength for ridge-type roughness, considered in \citet{zampino2022}. Given the striking similarity with experimental and numerical observations, this reinforces the idea of secondary motions as the output response of the linearised Navier-Stokes operator, because the response does not strongly depend on the surface features, but rather on the nature of the operator and the forcing spanwise length scale.
 
% There is one prediction that . This concerns the occurrence of tertiary structures and flow reversal for specific surface configurations where the strips do not have equal width

From a practical standpoint, the advantage of the present approach is, undoubtedly, its computational efficiency. However, the ability to rapidly probe the parameter space has enabled progress to be made on a more fundamental standpoint. Specifically, previous work has suggested that secondary currents may be the time-averaged picture of naturally-occurring large scale motions locked in place by the heterogeneity. The robustness of the above-mentioned similarities between the flow structure predicted by the present framework and previous observations suggests an alternative input-output perspective whereby these currents are the output response of the Navier-Stokes operator linearised about the turbulent mean and subjected of a steady, streamwise-independent forcing localised at the wall, associated to the lateral perturbation of the surface characteristics. This perspective complements the well accepted viewpoint that instantaneous coherent structures in wall-bounded turbulence may be described to a satisfactory degree by the output properties of the linearised operator subjected to a random forcing \citep{McKEON_SHARMA_2010, hwangcossu2010}. Admittedly, this viewpoint does not explain the observed interplay between large scale motions and secondary structures \citep{wangsawijaya2022}, whereby energy of the former leeches into the latter. One possible explanation that is worth exploring further is that the mean flow distortion produced by the secondary currents may locally alter the selective amplification properties of the linearised operator, producing a spanwise modulation of the nature of large-scale motions that may be interpreted as an energy interaction between secondary currents and large-scale motions.

A second key output of this study is that it offers a unified perspective to examine both ridge-type and strip-type roughness. Examination of these two cases within the present framework shows that secondary motions produced by complex surface heterogeneities, e.g.~arbitrary combinations of elevation and roughness properties, may be seen as originating from two separate source mechanisms. The first is a lateral variation of the virtual origin from which the mean turbulence structure develops. The sign of this variation is opposite for ridge-type and strip-type roughness: the virtual origin is shifted upwards by ridges, and downwards by higher roughness. The second source mechanism is mediated by the lateral variation of the streamwise slip velocity in the vicinity of the wall, and the associated spanwise gradients of the streamwise velocity. This source mechanism captures the acceleration/deceleration perceived by the bulk flow above the troughs and crests of a non-planar topography, respectively, or when the mean roughness height varies laterally. In ridge-type roughness, we have shown that this source mechanism damps the first so that the resulting secondary motions are weaker compared to those that would be predicted from the first mechanism. In other words, \emph{for the same lateral variation of the virtual origin} strip-type roughness produces more intense secondary flows than ridge-type roughness. The caveat is that this perspective applies, in the limit where the lateral variation of surface attributes is ``small'', in the region of validity of the linear model. For finite amplitude perturbations of the surface attributes, these predictions would need to be further verified. In any case, the present modelling framework suggests that the mean roughness height, a geometric quantity used in previous studies that considered the coupling of roughness and elevation is not the important quantity to be monitored when investigating combinations of elevation and roughness. Instead, we suggest that the notion of the virtual origin, a dynamical parameter associated to the downward shift of the logarithmic distribution, should be considered.

Like all other analyses of the linearised Navier-Stokes operator, the present approach yields useful insight but has its limitations. The ability of a one-equation turbulence {model} equipped with a nonlinear Reynolds stress model to capture the unsteady motion of secondary {structures} may be questioned. A more extensive {assessment of alternative turbulence modelling strategies and a comparison with high-fidelity simulations is certainly warranted in future work, although it must be pointed out that the SA-QCR model used here} does seem to capture fairly well the anisotropic nature of the Reynolds stress tensor in square-duct flow \citep{modesti2020priori}. {In this regard, it is speculated that such alternatives may not necessarily lead to significant qualitative changes in the predictions obtained here. In fact, we argue that the selectivity of the linearised Navier-Stokes operator may be dominant, with minimal effects of the turbulence modelling strategy adopted. This speculation is supported by the extensive evidence that useful predictions have been made using linearised Navier-Stokes equation approaches using much simpler turbulence modelling techniques than used here. Such a speculation may be verified simply by examining the response obtained from other turbulence modelling strategies. Given that linearisation of complex models is a tedious task, nonlinear calculations using existing solvers may be performed for sufficiently small roughness variations that the linear approximation is reasonably valid.} {A further limitation is that it is unclear how sensible is the streamwise independence assumption used here when secondary motions display a strong meandering behaviour \citep{kevin2019}.}

{Further, the importance of nonlinear effects neglected in the present linear framework in altering the structure of secondary flows is not clearly understood. Convergence of the expansion \eqref{lineardecomposition} for finite amplitude surface perturbations should be examined}. Analyses of the streamwise vorticity budget for ridge-type roughness \citep{castro2024} shows that nonlinear effects appear significant near sharp geometric features, and similar conclusions might apply to regions where roughness properties vary sharply. Nonlinear effects must also be introduced in order to predict the change in drag {\citep{ZampinoPhD}}, since the {one further limitation} of the present linear approach is that it offers no insight into the dependence of drag on the surface properties. This would be highly useful for predicting the drag of real-world surfaces, which is currently a topic of active research. {Clearly, the question is whether classical turbulence models and established rough-wall treatment strategies may adequately capture the drag characteristics of heterogeneous surfaces. A possible avenue forward would be to by-pass these modelling strategies and the small-amplitude assumption and instead adopt the framework reviewed in \citet{Zare.2020} to model the second-order statistics of the velocity fluctuations from the response of the linearised Navier-Stokes subjected to a structured stochastic forcing. However, the extension of this framework to rough surfaces would need to be considered first.}

% Finally, there are outstanding questions in the literature that may be worth

% Linear stability may be attempted to predict the longitudinal wavenumber at which a sinuous instability may appear, similarly. The base flow is easy to generate and we could show how the growth rate varies as a function of the strip width $S$ and of the streamwise wavenumber. cite paper that did it for the streaks. Alternatively it might be possible that the nonlinear travelling wave solutions of the URANS equations may be found \citep{jeong97, cossu2017}.

% - virtual origin of the turbulent viscosity profile which controls turbulent stresses.

% - two pathways that capture elevation and roughness

% - damping effect of streamwise velocity slip, so that topographies appear less effective in producing secondary flows per unit displacement of the virtual origin.

% - model the perturbation of the viscosity, so that only the orr sommerfield squire

\noindent
\textbf{Data access statement}: All data supporting this study will be openly available from the University of Southampton repository.\\

\noindent
\textbf{Acknowledgments}: We acknowledge funding from EPSRC (Grant ref: EP/V00199X/1). We are grateful for the insightful discussions with Prof Ian Castro and Prof Jae-Wook Kim.

\noindent
\textbf{Declaration of Interests}: The authors report no conflict of interest.\\

\appendix
\section{{Linearization of the normalised rotation tensor}}\label{appendix:qcr}
{The normalised rotation tensors at order zero and order one are
 \begin{align}
    O^{(0)}&=\left[\begin{matrix} 0 & sign(\Gamma) & 0 \\
    -sign(\Gamma) & 0& 0\\
    0 &0 &0	
    \end{matrix}\right], \\
   O^{(1)}&=\left[\begin{matrix} 0 & 0 & \displaystyle\frac{sign(\Gamma)}{\Gamma} \frac{\partial u_{1}^{(1)}}{\partial x_{3}} \\
    0 & 0& \displaystyle \frac{sign(\Gamma)}{\Gamma} \left(\frac{\partial u_{2}^{(1)}}{\partial x_{3}}-\frac{\partial u_{3}^{(1)}}{\partial x_{2}} \right)\\ \displaystyle
    -\frac{sign(\Gamma)}{\Gamma} \frac{\partial u_{1}^{(1)}}{\partial x_{3}}& \displaystyle -\frac{sign(\Gamma)}{\Gamma} \left(\frac{\partial u_{2}^{(1)}}{\partial x_{3}}-\frac{\partial u_{3}^{(1)}}{\partial x_{2}} \right) &0	
    \end{matrix}\right],
   \end{align}
where $\Gamma$ is the zero-order wall-normal gradient of the streamwise velocity and $sign$ is the sign function.}

\bibliographystyle{jfm}
\bibliography{biblio}

\end{document}